\documentclass[referee]{raa}    
\usepackage{newtxmath}
\usepackage{graphicx,times}             
\usepackage{natbib}
\usepackage{amsmath}

\usepackage{amssymb}
\usepackage{url}
\usepackage{float}
\usepackage{textcomp}
\usepackage{url}
\usepackage{pdfpages}
\usepackage{tabularx}
\usepackage{makecell}
\usepackage{upgreek}
\usepackage{threeparttable}

\bibpunct{(}{)}{;}{a}{}{,}

\begin{document}

  \title{Deep Learning Applications Based on WISE Infrared Data: Classification of Stars, Galaxies and Quasars}

   \volnopage{Vol.0 (20xx) No.0, 000--000}      
   \setcounter{page}{1}          

   \author{Guiyu Zhao 
      \inst{1}
   \and Bo Qiu
      \inst{1}\thanks{corresponding author}
   \and A-Li Luo
      \inst{2,3,}\thanks{corresponding author}
   \and Xiaoyu Guo
      \inst{1}
   \and Lin Yao
      \inst{1}
   \and Kun Wang
      \inst{1}
   \and Yuanbo Liu
      \inst{1}
   }

   \institute{School of Electronic and Information Engineering,Hebei University of Technology, Tianjin, 300401, China; {\it 1263730840@qq.com, qiubo@hebut.edu.cn, 1799507446@qq.com, 1286789387@qq.com, 1848896968@qq.com, 1220617881@qq.com}\\
        \and
             CAS Key Laboratory of Optical Astronomy, National Astronomical Observatories, Beijing 100101, China; {\it lal@bao.ac.cn}\\
        \and
             University of Chinese Academy of Sciences, Beijing 100049, China\\
\vs\no
   {\small Received 2023 March 2; accepted 2023 May 14}}

\abstract{ The Wide-field Infrared Survey Explorer (WISE) has detected hundreds of millions of sources over the entire sky. However, classifying them reliably is a great challenge due to degeneracies in WISE multicolor space and low detection levels in its two longest-wavelength bandpasses. In this paper, the deep learning classification network, IICnet (Infrared Image Classification network), is designed to classify sources from WISE images to achieve a more accurate classification goal.
IICnet shows good ability on the feature extraction of the WISE sources.
Experiments demonstrates that the classification results of IICnet are superior to some other methods; it has obtained 96.2$\%$ accuracy for galaxies, 97.9$\%$ accuracy for quasars, and 96.4$\%$ accuracy for stars, and the Area Under Curve (AUC) of the IICnet classifier can reach more than 99$\%$.
In addition, the superiority of IICnet in processing infrared images has been demonstrated in the comparisons with VGG16, GoogleNet, ResNet34, MobileNet, EfficientNetV2, and RepVGG—fewer parameters and faster inference. The above proves that IICnet is an effective method to classify infrared sources.
}

   \authorrunning{Guiyu Zhao et al.}            
   \titlerunning{Classification of Stars, Galaxies and Quasars}  

   \maketitle

%
%
\section{Introduction}           
\label{sect:Introduction}

Infrared astronomical observation is one of the most important branches of observational astronomy today, which mainly focused on the study of various types of celestial sources in the universe through observations in the infrared band \citep{Glass}, and the objects which are too dim in the visible band can also be detected in the infrared band.

The Earth is surrounded by a thick layer of atmosphere that contains many substances, such as water vapor, carbon dioxide, oxygen, and ozone. They have a strong scattering and absorption effect on celestial radiation from outer space at infrared wavelengths \citep{Liou}, which limits ground-based infrared astronomical observations.
Some initial observatories, such as the Kuiper Airborne Observatory (KAO) \citep{Erickson} and Stratospheric Observatory for Infrared Astronomy (SOFIA) \citep{EricksonEF}, developed to infrared space telescopes, such as the Infrared Astronomical Satellite (IRAS) \citep{Duxbury}, the Infrared Space Observatory (ISO) \citep{Kessler}, and the Wide-field Infrared Survey Explorer (WISE) \citep{Wright}.

Classification is an essential means for humans to acquire knowledge, and the problem of classifying celestial targets has been studied for a long time \citep{lintott2008galaxy}. The classification scheme of galaxies, quasars, and stars is one of the most fundamental classification tasks in astronomy \citep{kim2016star, ethiraj2022classification}. The classification of celestial objects usually includes spectral classification and morphological image classification.

Spectral classification is very popular and there are a lot of reported works. The classification of stars, galaxies, and quasars by spectroscopy has been studied commonly, but generally it requires a large workload by comparing the observed spectra with a template.
Later, a random forest method was also used to do the same task, but the classification accuracy of quasars was only 94$\%$ \citep{Bai}.

The morphological classification is also a common experiment.
A self-supervised learning method was used to classify the 3 classes based on photometric images, and the accuracy could only reach 88$\%$ \citep{Martinazzo}. Some researchers have classified sources into stars, galaxies, and quasars with high accuracy based on SDSS photometric images using deep learning methods, which is instructive for our work \citep{HeZhendong}.

A support vector machine (SVM) \citep{Steinwart} method was used to classify 3 classes based on WISE and SDSS with information from the W1 band \citep{Kurcz}.\add{ }Classification of galaxy morphology based on WISE infrared images has been previously investigated \citep{GuoXiaoyu}, and we have taken the classification of infrared images a step further.

In this paper, the data used with their pre-processing details are introduced in Section~\ref{sec2}; the Infrared Image Classification Network (IICnet) with the modules is introduced described in Section~\ref{sec3}; the classification results are presented, and some comparison experiments are performed in Section~\ref{sec4}; the experimental results are analyzed in Section~\ref{sec5}; and the summary in Section~\ref{sec6}.

\begin{figure*}
	\centering
	\subfloat[A star image of 600 arcsec]{\includegraphics[height=4cm,width=4cm]{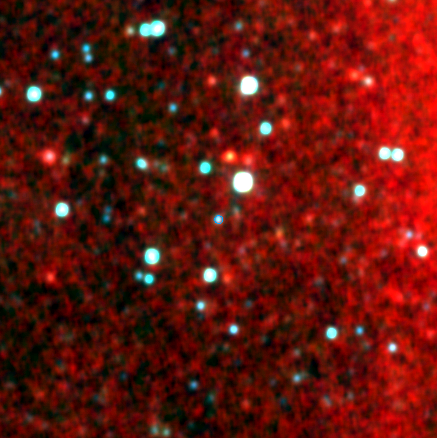}\label{fig1: (a)}}
	\quad
	\subfloat[A star image of 50 arcsec]{\includegraphics[height=4cm,width=4cm]{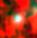}\label{fig1: (b)}}
	\caption{Images corresponding to different arc seconds. We chose 50 arcsec for processing, as WISE website defaults to 600 arcsec.}
	\label{fig1}
\end{figure*}

\section{DATA}
\label{sec2}
The dataset is constructed on some selected infrared image data from WISE\footnote{\url{https://irsa.ipac.caltech.edu/applications/wise/}}.

\subsection{Data preparation}
\label{sec2.1}
WISE has four bands, W1, W2, W3, and W4, at wavelengths of 3.4 $\upmu$m, 4.6 $\upmu$m, 12 $\upmu$m, and 22 $\upmu$m, respectively \citep{Wright}. The WISE all-sky images and source catalog, released in March 2012, contain over 563 million objects and provide a massive amount of information on MIR properties of many different types of celestial objects and their related phenomena \citep{Wright,Tu}.
By 2013, WISE had detected over 747 million objects with SNR \textgreater 5 and publicly released in the AllWISE source catalog \citep{Cutri}. 

When acquiring raw data in WISE, if the image size is set to 600 arcsec (the default value), there will be too many sources in the image, as shown in Fig.~\ref{fig1: (a)}. To find the specific source corresponding to the RA and Dec, the image size is set to 50 arcsec, as shown in Fig.~\ref{fig1: (b)}. The data corresponding to each RA and Dec in this paper was obtained in INFRARED SCIENCE ARCHIVE (IRSA)\footnote{\url{https://irsa.ipac.caltech.edu/frontpage/}}. The band information of W1, W2, W3, and W4 of the corresponding sources are obtained from WISE after the crossover between SDSS\footnote{\url{http://skyserver.sdss.org/CasJobs/SubmitJob.aspx}} and WISE to form the experimental database of this project.

\begin{figure*}
	\centering
	\includegraphics[scale=0.05]{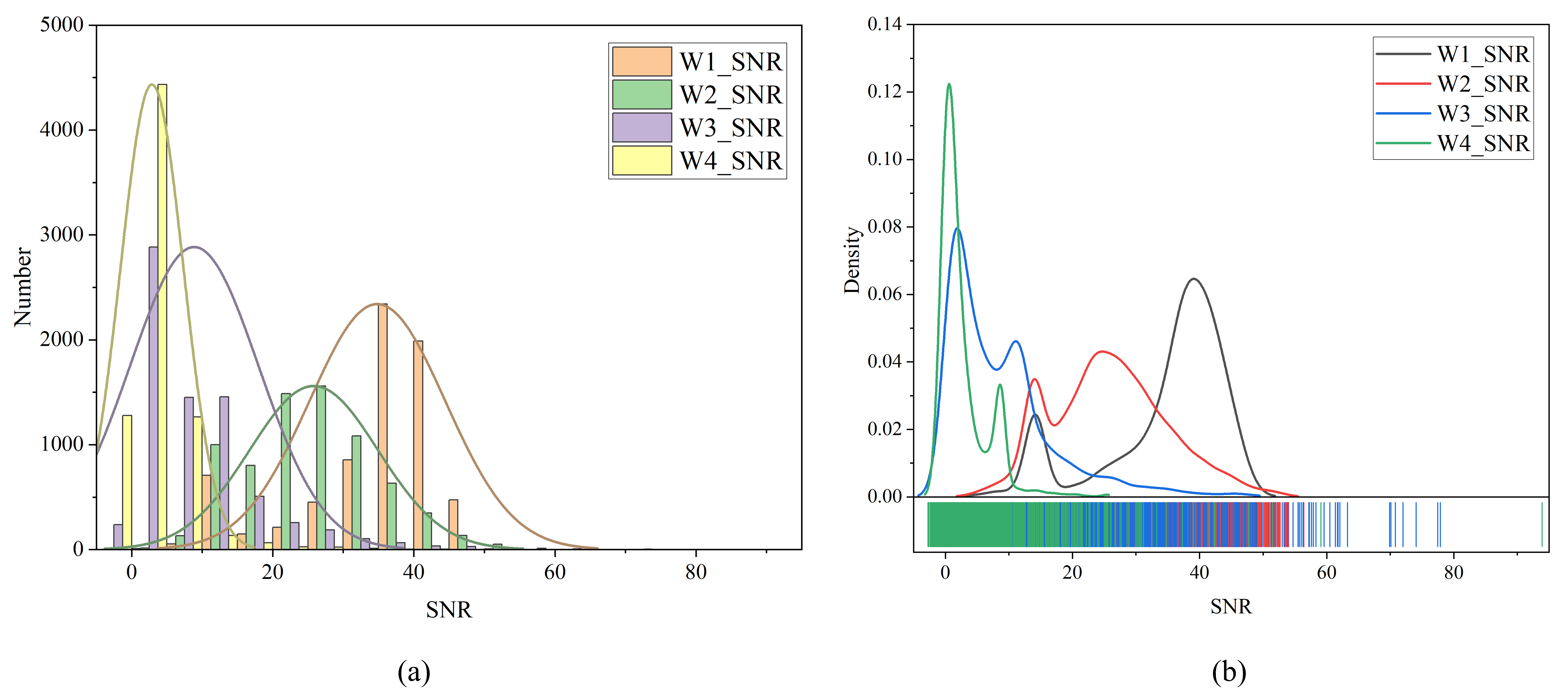}
	\caption{Statistical and probability distribution figures of SNR for the four bands. (a) Statistical figure of SNR. (b) Probability distribution of SNR.}
	\label{fig2}
\end{figure*}

\begin{figure}[h]
	\centering
	\includegraphics[scale=0.22]{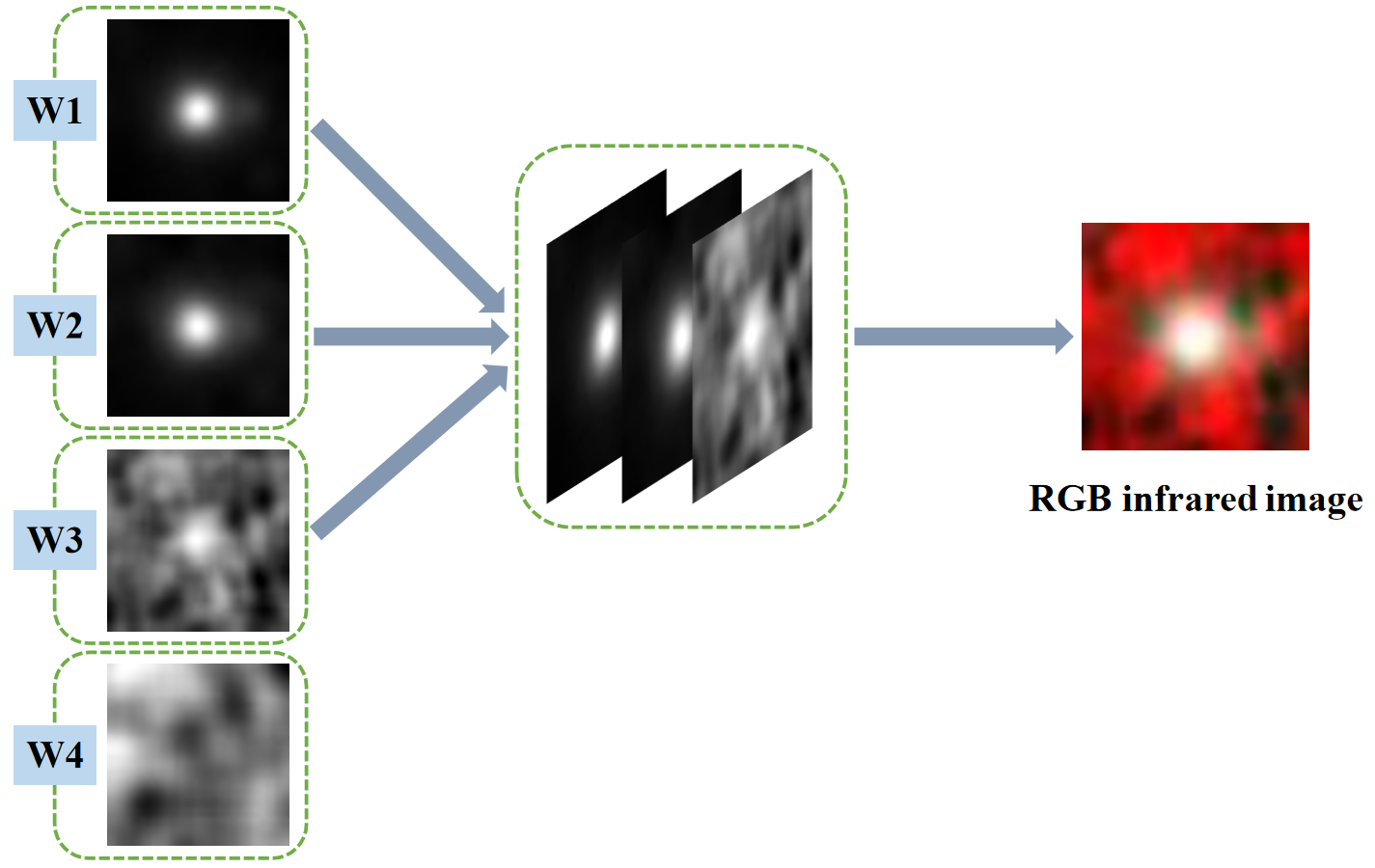}
	\caption{A galaxy image of W1, W2, W3, W4 bands and a RGB infrared image synthesized by W1, W2, W3.}
	\label{fig3}
\end{figure}

\subsection{Image pre-processing}
\label{sec2.2}
WISE image classification can be adversely affected by excessive dust around the sources, and the presence of more dust in the W4 band and the lower signal-to-noise ratio (SNR) compared to the other three bands are shown in Fig.~\ref{fig2}. W4 exhibits a significantly lower SNR than the other three bands, therefore, in this paper, the W1, W2, and W3 bands have been used as the three channels of the RGB image to synthesize the infrared image, as shown in Fig.~\ref{fig3}.

Furthur more, 7298 galaxy images, 7215 quasar images, and 7223 star images are chosen to form the dataset finally. Their numbers are approximately equal to each other to ensure data balance between different classes for satisfying the demands of deep learning algorithms. The dataset is randomely divided into training, validation, and test sets with a ratio of 8:1:1, as shown in Table~\ref{tab1}.

\begin{table}[t]
	\centering
	\caption{Datasets division of three types of celestial bodies.}
	\label{tab1}
	\begin{tabular}{lccc} 
		\hline
		Type & Training sets & Test sets & Validation sets\\
		\hline
		Galaxy & 5838 & 730 & 730\\
		Quasar & 5772 & 722 & 721\\
		Star   & 5778 & 723 & 722\\
		\hline
	\end{tabular}
	\begin{tablenotes}
		\item\textit{Note}. Some celestial objects in this work have corresponding spectra to ensure their type being determined.
	\end{tablenotes}
\end{table}

\begin{figure}
	\centering
	\subfloat[]{\includegraphics[height=2.6cm,width=2.6cm]{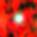}}
	\quad
	\subfloat[]{\includegraphics[height=2.6cm,width=2.6cm]{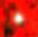}}
	\quad
	\subfloat[]{\includegraphics[height=2.6cm,width=2.6cm]{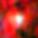}}
	\caption{Sample images for each type. The three types of objects have confusing features. (a) A galaxy. (b) A star. (c) A quasar.}
	\label{fig4}
\end{figure}

One of the difficulties of the classification is that some infrared images of galaxies, quasars, and stars look highly similar. As shown in Fig.~\ref{fig4}, they all have a brighter light source in the image center and lack obvious image features that can clearly distinguish them from each other by human eyes.
This paper introduces the IICnet method to do the classification automatically. The basis of this method is that convolutional neural networks can extract image features that human eyes cannot distinguish \citep{Egmont}.

\begin{figure}
	\centering
	\includegraphics[scale=0.4]{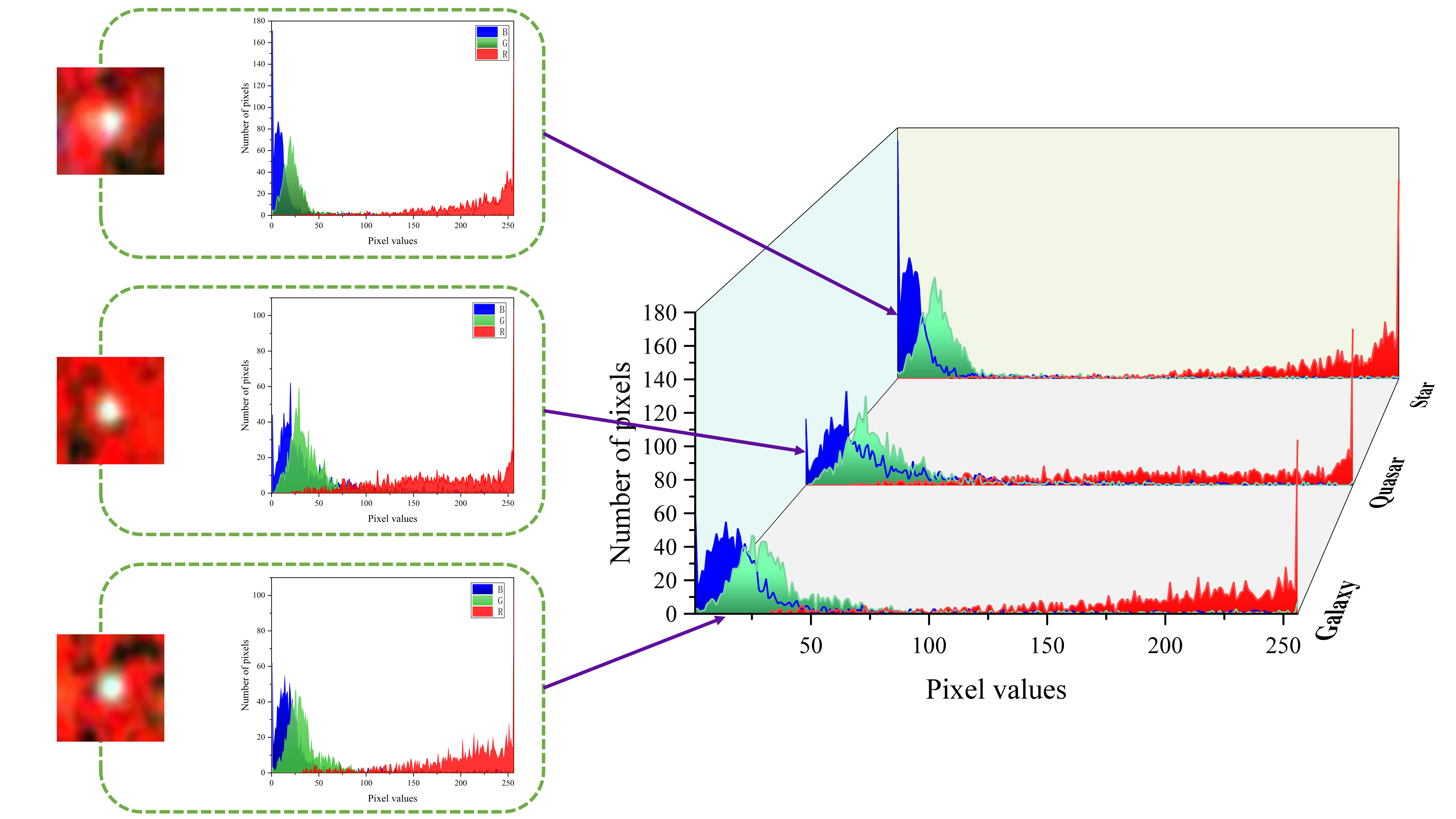}
	\caption{3D waterfall of galaxies, stars and quasars. On the left there are RGB histograms of sample images of a star, a quasar, and a galaxy, respectively, and on the right it is a 3D waterfall combination of the left.}
	\label{fig5}
\end{figure}

When the RGB histogram is used to distinguish the three images in Fig.~\ref{fig4}, the results are shown in Fig.~\ref{fig5}. It can be found that 3 histograms are similar to each other. So simple image features like histogram cannot distinguish the 3 types, the deep learning method is designed to do the classification.

\begin{figure}[h]
	\centering
	\includegraphics[scale=0.7]{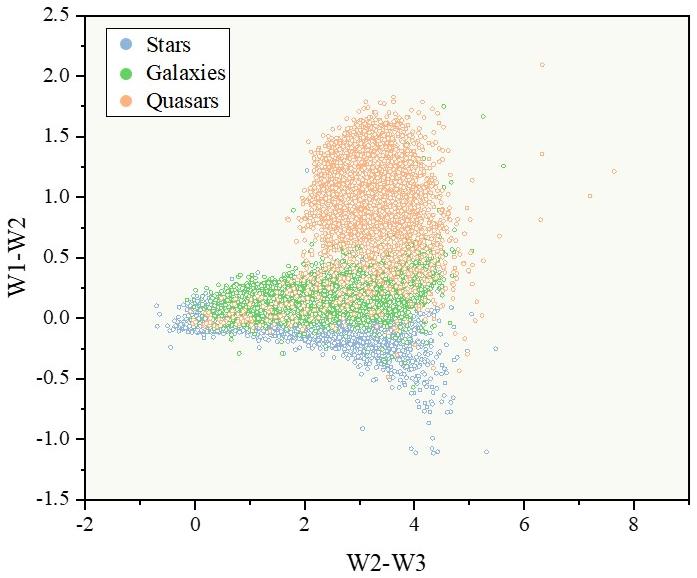}
	\caption{Color–color diagram showing the locations of 3 types. There are large areas of overlap between the three types of objects.}
	\label{fig6}
\end{figure}

In the low-redshift universe, the stars and galaxies of W1-W2 exhibit very similar colors \citep{Kurcz}.
If the color-color diagram composed of W1, W2, and W3 is used to analyze the distributions between stars, galaxies, and quasars \citep{Wright} (Fig.~\ref{fig6}), it can be found that there are large overlap regions among the three types, especially the overlap between stars and galaxies is very obvious.
This illustrates that it is difficult to accomplish the infrared image classification task by conventional means.

\begin{figure}
	\centering
	\includegraphics[width=\linewidth]{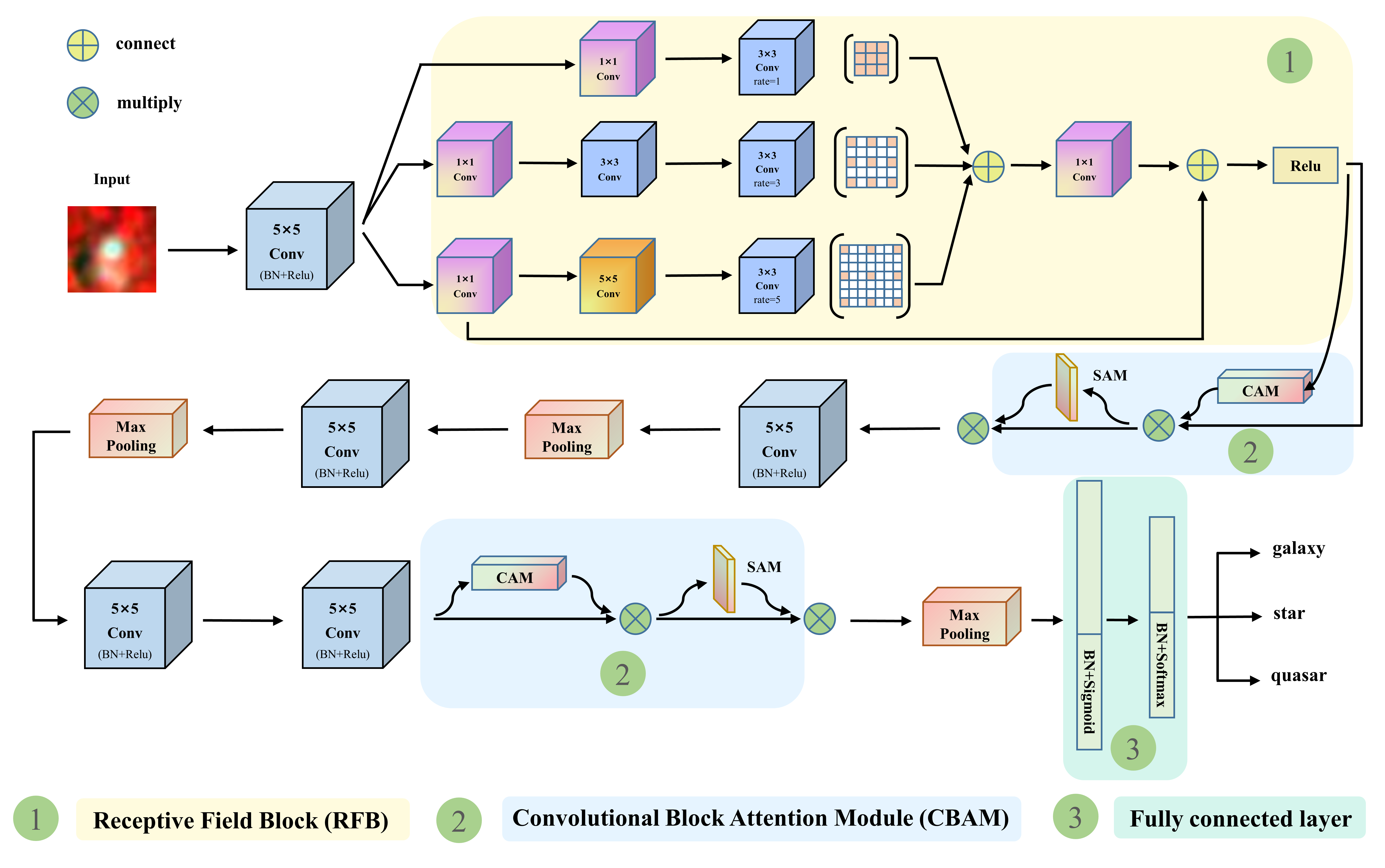}
	\caption{Network structure of IICNet. This network is used for the classification of infrared images. The input is a 3-channel matrix with a size of 80×80 and contains data in the three bands of W1, W2 and W3, and the output is the type of input source predicted by the network.}
	\label{fig7}
\end{figure}

\begin{figure}
	\centering
	\includegraphics[scale=0.1]{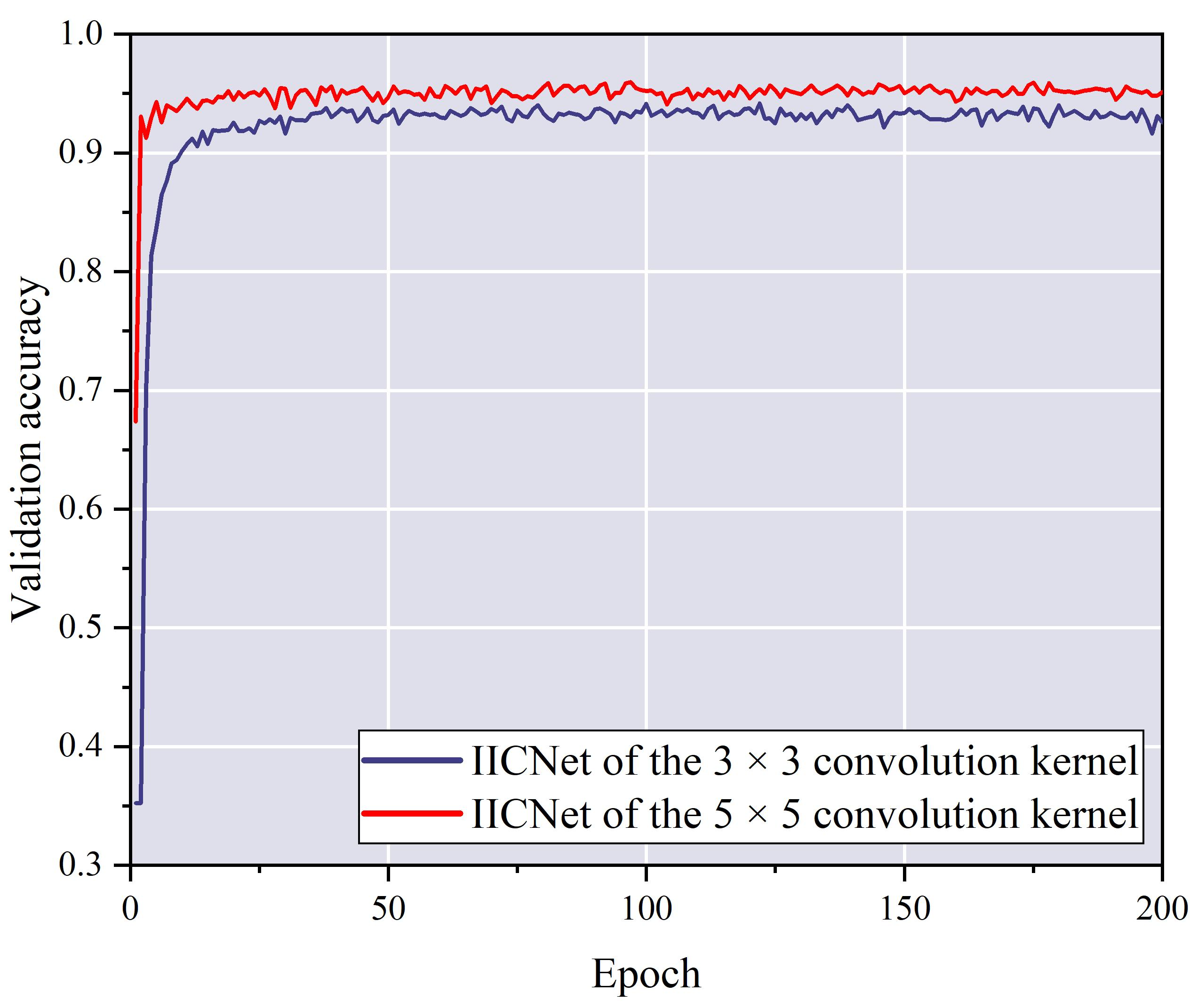}
	\caption{Verification accuracy of different convolution kernels. The accuracy of the 5$\times$5 convolution kernel is significantly higher than that of the 3$\times$3.}
	\label{fig8}
\end{figure}

\begin{figure}
	\centering
	\includegraphics[scale=0.05]{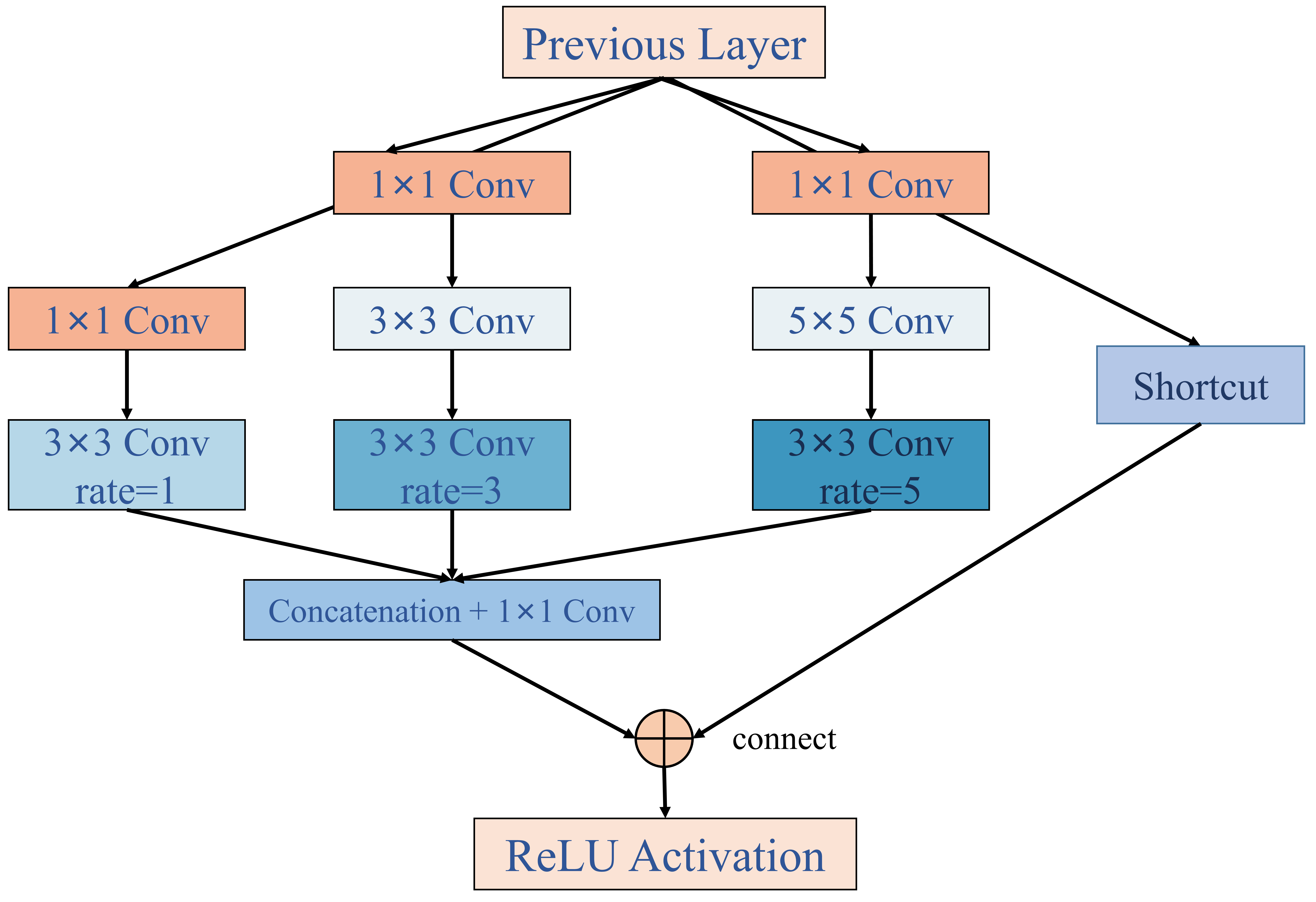}
	\caption{The architectures of RFB.}
	\label{fig9}
\end{figure}

\begin{figure}
	\centering
	\subfloat[Channel attention module (CAM)]{\includegraphics[scale=0.05]{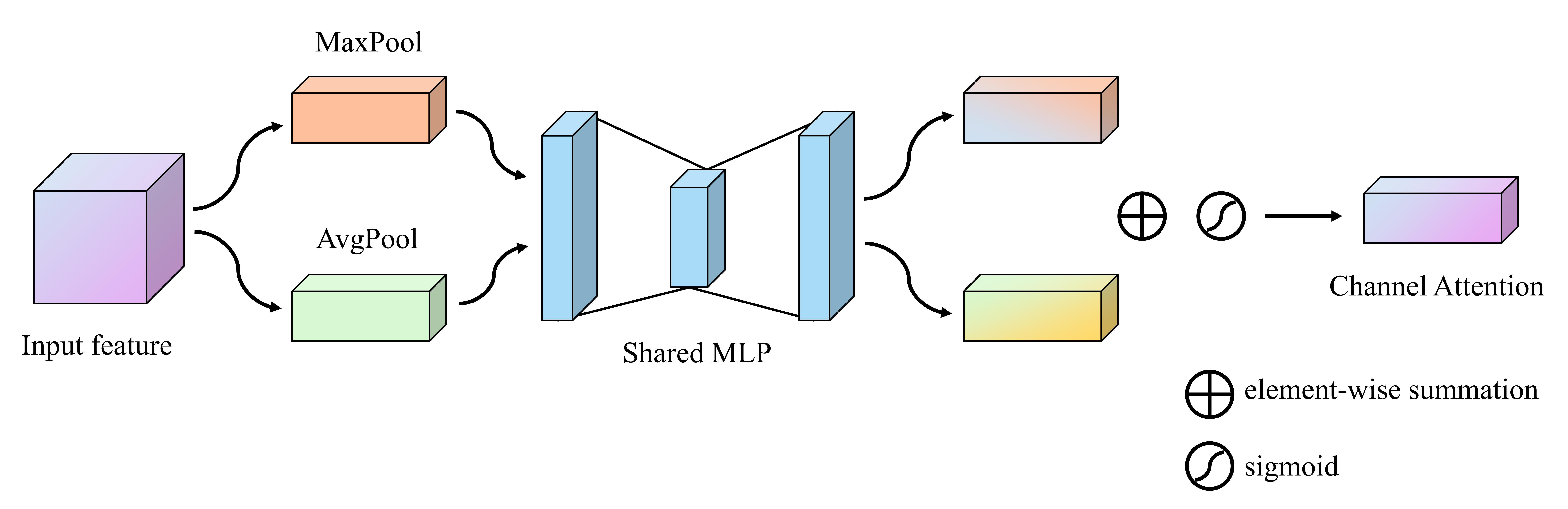}\label{fig10: a}}
	\quad
	\subfloat[Spatial attention module (SAM)]{\includegraphics[scale=0.05]{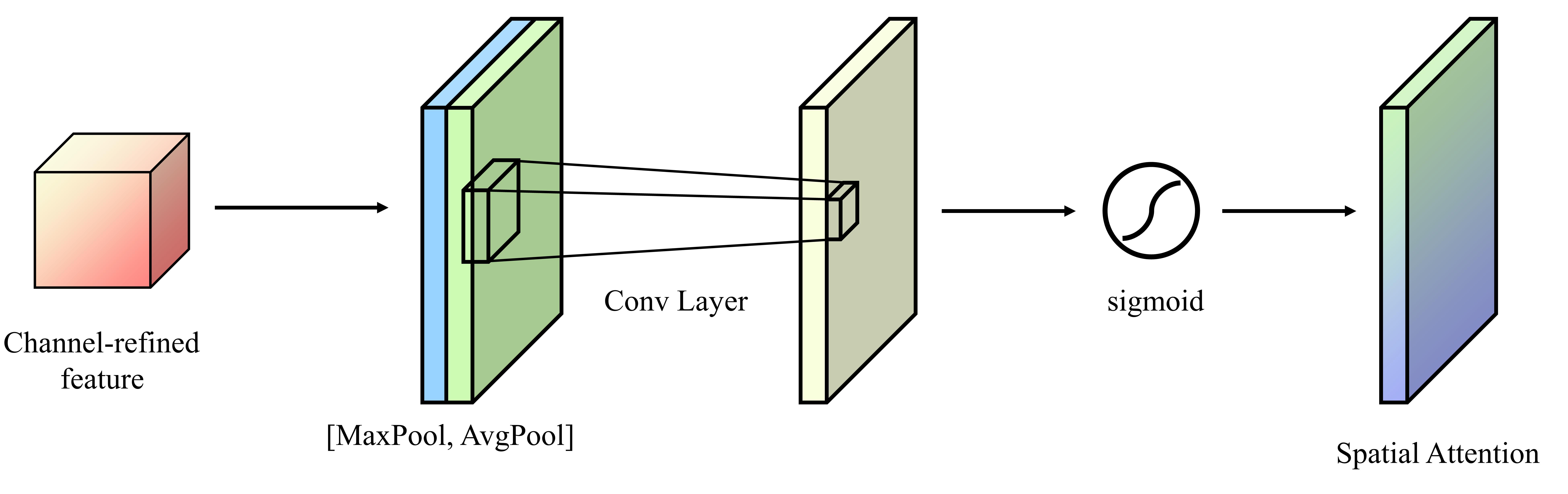}\label{fig10: b}}
	\caption{Diagram of each attention sub-module. CAM makes use of average and maximum pooling in simultaneously. SAM connects two feature layers together to create one feature layer.}
	\label{fig10}
\end{figure}

\section{METHODS}
\label{sec3}
In this paper, a new deep learning algorithm IICnet is designed to accomplish the task of infrared image classification.
For this task, experiments are conducted based on the Pytorch architecture and the Python programming language.
An NVIDIA TESLA V100 GPU (5120 CUDA cores and 32GB of video memory) is used for training.
\subsection{Infrared image classification network: IICnet}
\label{sec3.1}

The structure of IICnet is shown in Fig.~\ref{fig7}. The network includes five convolutional layers, three down-sampling layers (pooling layers), one feature extraction module (Receptive Field Block, RFB) \citep{RFB}, and two convolutional block attention modules (CBAM) \citep{CBAM} at the beginning and the end.

In IICnet, the first block is a large convolutional kernel of 5$\times$5, it has been demonstrated by several researchers that large convolutional kernels are more capable of extracting semantic information \citep{largekernel}. It extracts information from an image's more extensive neighborhood range to ensure its relative integrity after it starts convolution.
The subsequent addition of the BN layer and ReLU can suppress gradient explosions and help extract deeper semantic information.
The experiments demonstrate that the 5$\times$5 convolutional kernel for this task outperforms the 3$\times$3 kernel. As shown in Fig.~\ref{fig8}, the validation accuracy of the network using the 5$\times$5 convolutional kernel is significantly higher than that of the 3$\times$3 kernel.

After the first layer of convolution, the raw feature map is generated and in the following fed to the Receptive Field Block (RFB) (the first module) for furthur processing. As shown in Fig.~\ref{fig9}, RFB is a feature extraction module that can enhance the feature extraction capability of the network by simulating the perceptual field of human vision. The first half of the module is similar to GoogleNet in which it can simulate group receptive fields of various sizes and adds dilated convolution to increase the receptive fields effectively. The latter half reproduces the relationship between the size and eccentricity of the population receptive field (pRF) \citep{pRF} in the human visual system, increasing the distinguishability and robustness of the features.

An attention module, the Convolutional Block Attention Module (CBAM) (the second module), is connected after the RFB and at the last layer of the network, respectively. CBAM not only indicates the direction of attention but also improves the representation of regions of interest. IICnet aims to improve feature representation by focusing on essential features and suppressing unnecessary ones. The channel and spatial attention modules are combined by CBAM, as shown in Fig.~\ref{fig10}. The Channel Attention Module (CAM) is shown in Fig.~\ref{fig10: a}. After the feature map is input, the one-dimensional vector of channel attention is first obtained through the global MaxPool and the global AvgPool; the respective one-bit vector is obtained after a shared Multi-Layer Perception (MLP) for element addition. Finally the spatial attention vector is obtained through sigmoid activation. Through the above process, the CAM can focus on the meaningful information in the image. The Spatial Attention Module (SAM) is shown in Fig.~\ref{fig10: b}, which is complementary to channel attention focuses on the target's location information. The Spatial SAM first uses MaxPool and AvgPool to obtain the channel-refined features in CAM, concatenates them and generates a feature descriptor, and finally activates them by sigmoid to obtain the feature map of SAM. The joint use of the two modules can achieve better results. The equations for CAM and SAM are expressed as follows:
\begin{equation}
	\begin{aligned}
		\mathbf{M}_{\mathbf{c}}(\mathbf{F}) &=\sigma(\textit{MLP}(\textit{AvgPool}(\mathbf{F}))+\textit{MLP}(\textit{MaxPool}(\mathbf{F}))) \\
		&=\sigma\left(\mathbf{W}_{\mathbf{1}}\left(\mathbf{W}_{\mathbf{0}}\left(\mathbf{F}_{\mathbf{avg}}^{\mathbf{c}}\right)\right)+\mathbf{W}_{\mathbf{1}}\left(\mathbf{W}_{\mathbf{0}}\left(\mathbf{F}_{\max }^{\mathbf{c}}\right)\right)\right)
	\end{aligned}
\end{equation}
where $\sigma$ denotes the sigmoid function. $\mathbf{W}_{\mathbf{0}}$ and $\mathbf{W}_{\mathbf{1}}$ are the weights of the MLP, and they are shared for both inputs and the ReLU activation function is followed by $\mathbf{W}_{\mathbf{0}}$. $\mathbf{F}_{\mathbf{avg}}^{\mathbf{c}}$ and $\mathbf{F}_{\max }^{\mathbf{c}}$ which denote average-pooled features and max-pooled features respectively.
\begin{equation}
	\begin{aligned}
		\mathbf{M}_{\mathbf{s}}(\mathbf{F}) &=\sigma\left(f^{7 \times 7}([\textit{AvgPool}(\mathbf{F}) ; \textit{MaxPool}(\mathbf{F})])\right) \\
		&=\sigma\left(f^{7 \times 7}\left(\left[\mathbf{F}_{\mathbf{a v g}}^{\mathbf{s}} ; \mathbf{F}_{\max }^{\mathbf{s}}\right]\right)\right)
	\end{aligned}
\end{equation}
where $\sigma$ denotes the sigmoid function and $f^{7 \times 7}$ represents a convolution operation with the filter size of 7$\times$7.

A softmax function is used at the end of the network to calculate the probability distribution of each class \citep{softmax}, which ultimately classifies the targets into stars, galaxies, and quasars.

The IICNet plays an essential role in improving the classification accuracy by performing feature extraction through each convolutional layer and downsampling layer. The RFB and CBAM modules can improve attention to the key position of the image, and the performance is significantly improved. Adam \citep{Adam} is one of the optimizers that uses hyperparameter computation efficiently, usually requires no tuning, and is simple to be implemented. It is used during training. In the training process, it is set to 200 epochs, and the initial value of the learning rate is set to $10^{-4}$, and after 50 epochs, it is set to half of the initial value ($5\times10^{-5}$), to ensure reasonable convergence of the training.

\subsection{Feature visualization of network layers}
\label{sec3.2}
\begin{figure}
	\centering
	\includegraphics[scale=0.05]{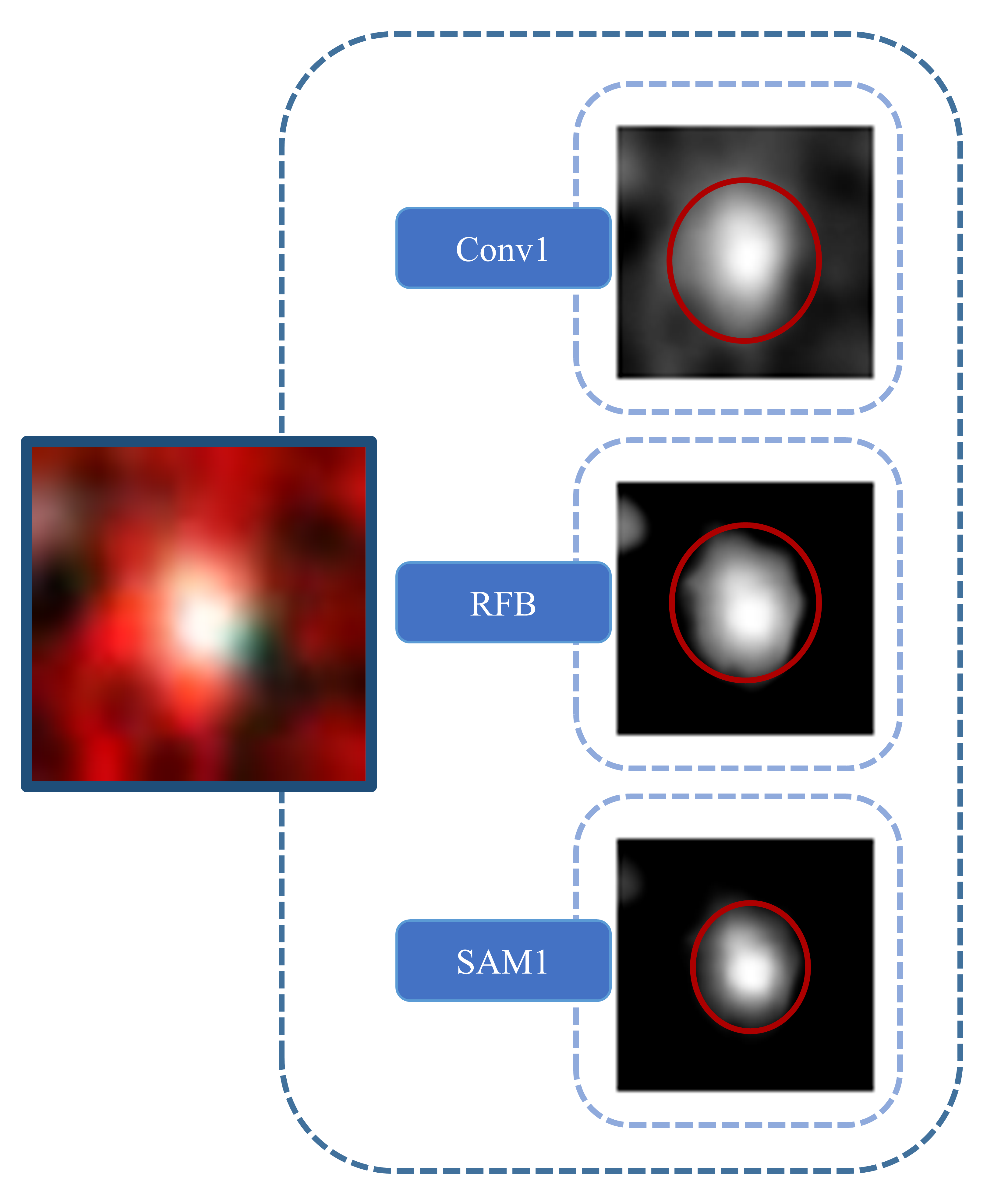}
	\caption{Middle layer visualization of the IICnet model. After the image passes through RFB and CBAM, the middle layer shows the focus on the central source.}
	\label{fig11}
\end{figure}

When analyzed with our dataset, the image central source is the most important part to be focused. There are different information around different sources, such as the predominance of red around stars, black and red around galaxies, and the more complex colors shown around quasars, with some blue and green mixed. The region of interest generated by IICnet can be observed by visualizing the features of the middle layer of the network, as shown in Fig.~\ref{fig11}. The feature maps are processed by the first layer convolution, RFB and CBAM respectively, and the regions of interest are more and more concentrated, which proves the importance of the feature extraction capability of RFB and the attention mechanism of CBAM for classification.

\begin{figure}
	\centering
	\includegraphics[scale=0.1]{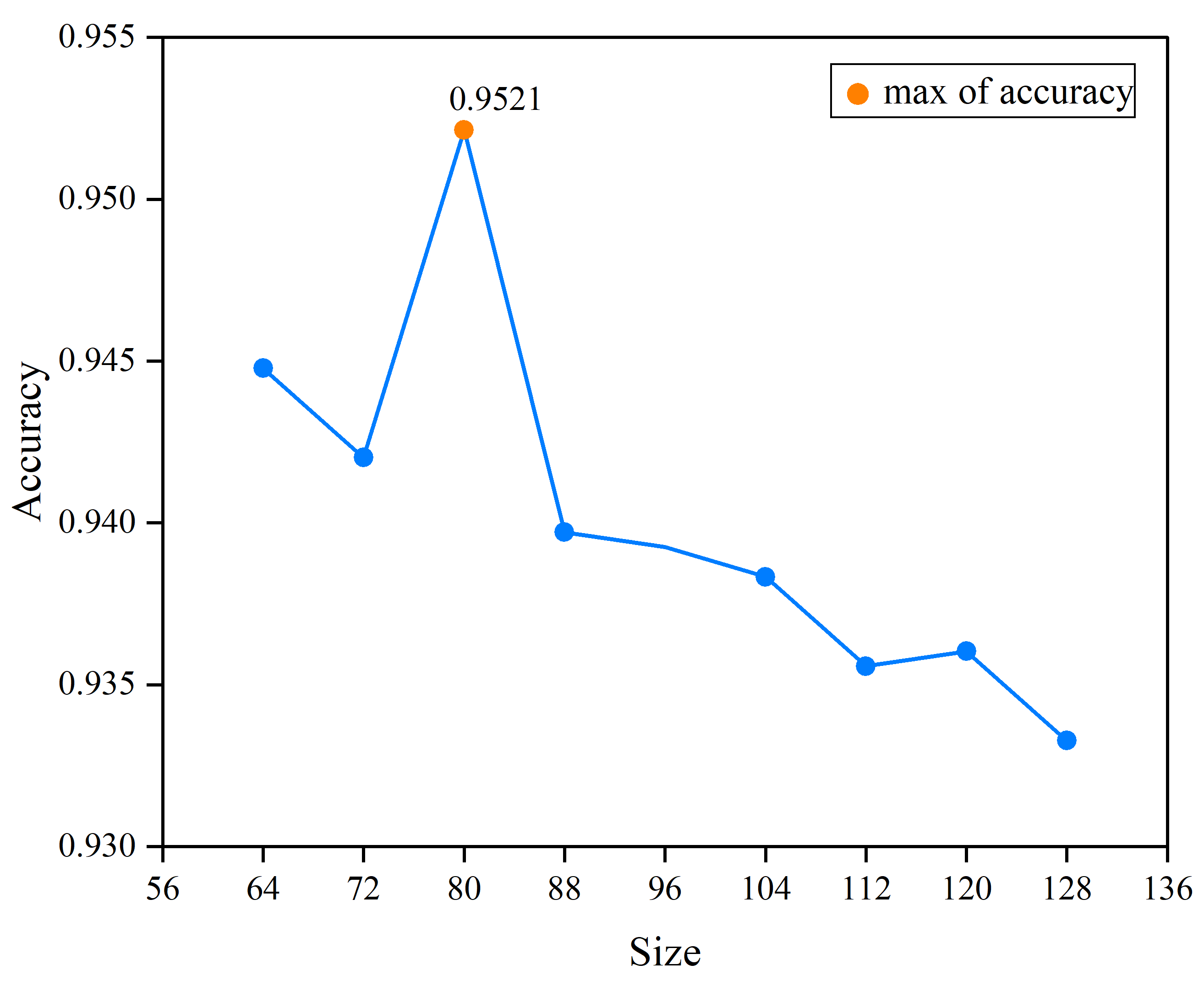}
	\caption{A plot of the relationship between input image size and IICnet accuracy. The accuracy achieves a maximum value of 0.9521 when the input image size is 80 × 80 pixels.} 
	\label{fig12}
\end{figure}

\begin{figure}
	\centering
	\subfloat[]{\includegraphics[scale=0.08]{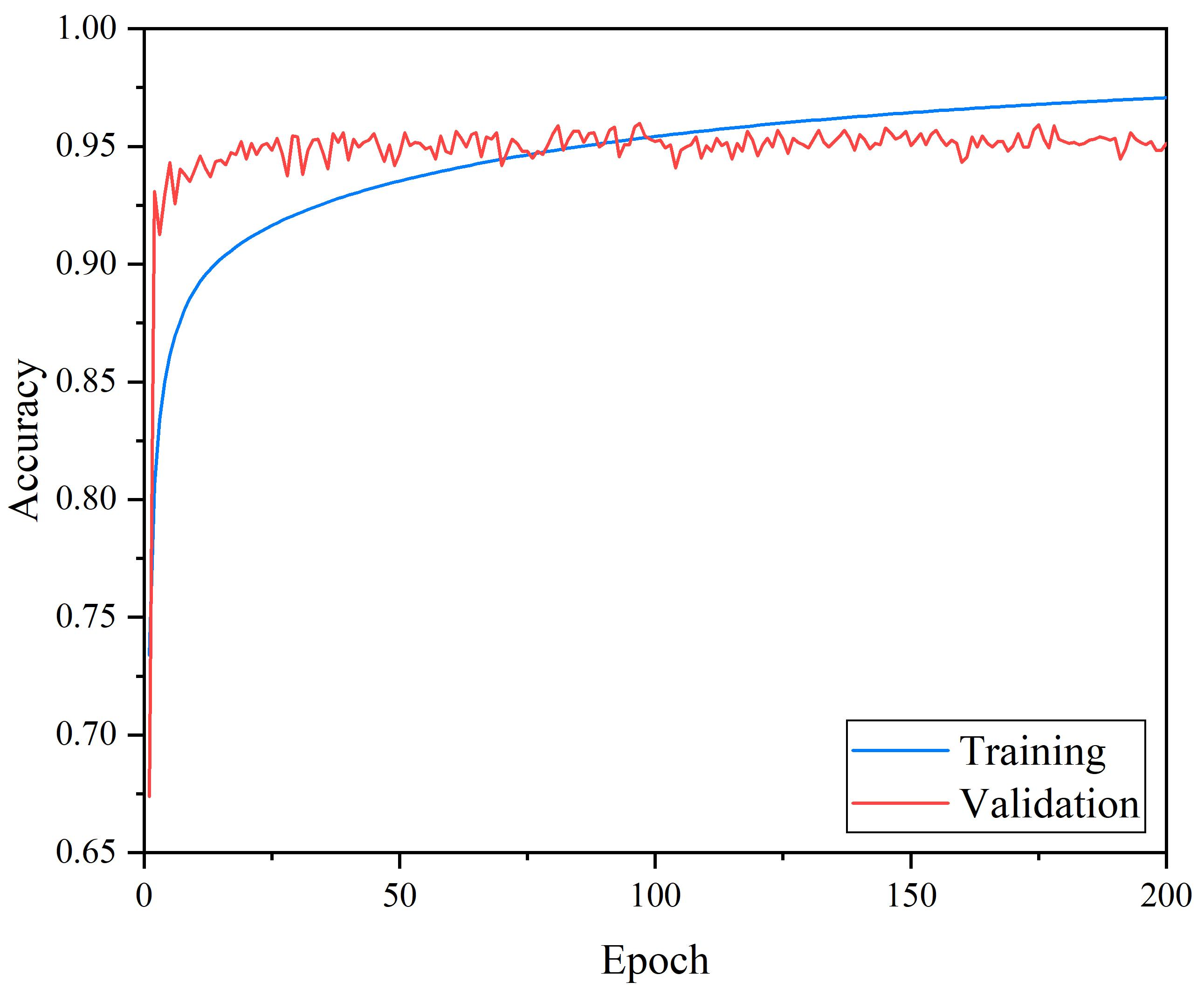}\label{fig13: a}}
	\quad
	\subfloat[]{\includegraphics[scale=0.08]{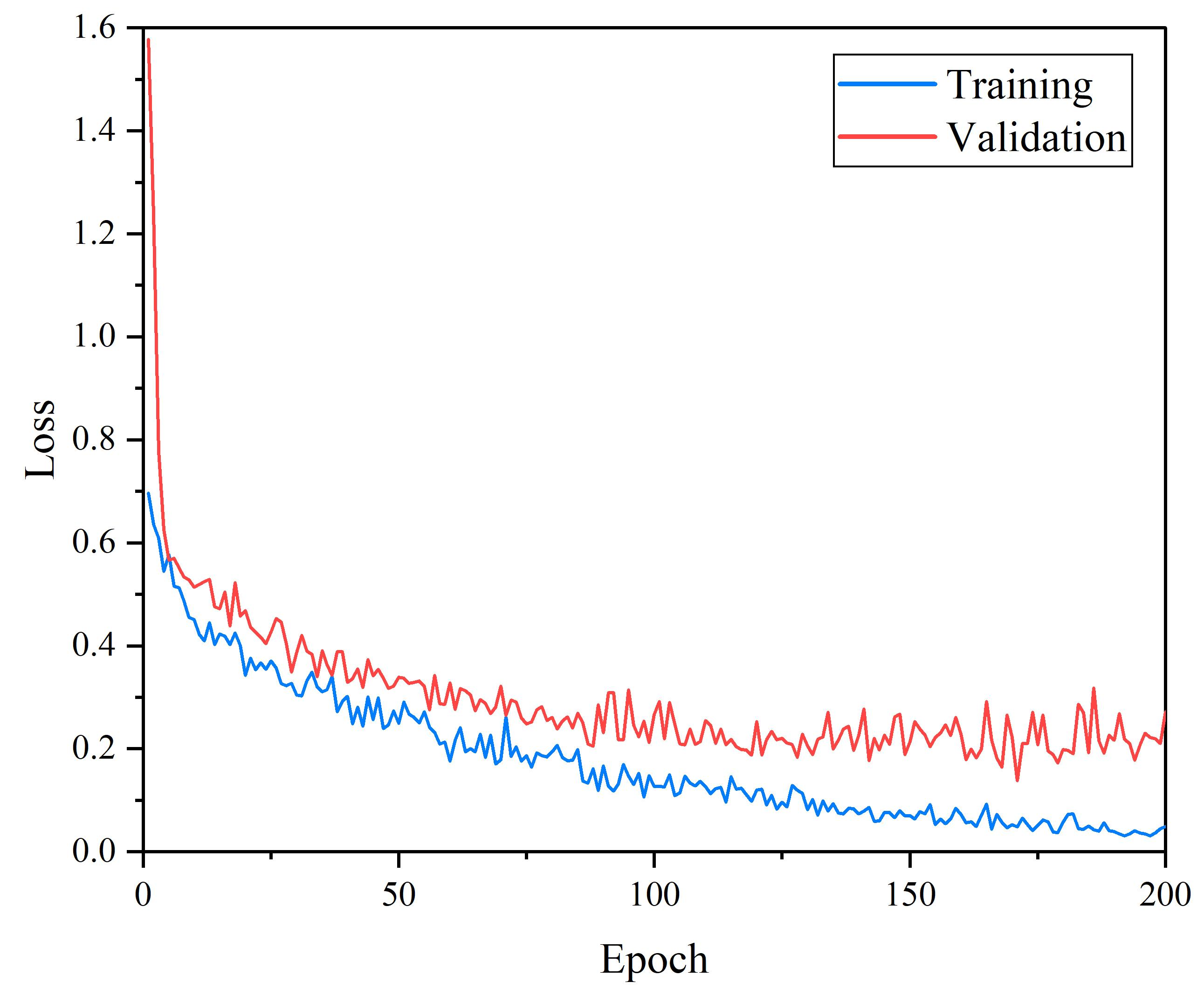}\label{fig13: b}}
	\caption{(a) The curve of IICnet’s loss against training set and validation set with epoch. (b) The curve of IICnet’s accuracy against training set and validation set with epoch.}
	\label{fig13}
\end{figure}

\section{RESULT}
\label{sec4}
\subsection{Influence of Image Size}
\label{sec4.1}
In the network of Convolutional Neural Network (CNN), the input image size is an essential factor affecting the the network's performance \citep{imagesize}. To obtain the optimal input size of the image, this paper tested the accuracy from 64$\times$64 to 128$\times$128, spanning 8, using 64$\times$64 as the starting size. The relationship between different input sizes and accuracy is shown in Fig.~\ref{fig12}. The image size achieves the highest accuracy at 80$\times$80. The accuracy gradually decreases as the image size increases, so 80$\times$80 is the most adaptable size for IICnet. 

\subsection{Influence of Epoch}
\label{sec4.2}
In this paper, the pre-processed infrared images of galaxies, stars, and quasars are input into IICnet, and the accuracy and loss obtained through the experiments are shown in Fig.~\ref{fig13}. In this experiment, accuracy and loss were analyzed through 200 epochs. The accuracy increased with the increase of epochs and then leveled off. The loss decreases as the epoch increases and then levels off. The accuracy of the validation set can reach up to 95$\%$ or more. IICnet's ability to get better results on infrared image classification is proven.

\subsection{Evaluation Indices}
\label{sec4.3}

\begin{table}
	\centering
	\caption{The classification index of IICnet including Precision, Recall, Specificity, F1-score, and Accuracy.}
	\label{tab2}
	\begin{tabular}{lccccc} 
		\hline
		Type & Precision & Recall & Specificity & F1-score & Accuracy \\
		\hline
		Galaxy & 95.4$\%$ & 93.2$\%$ & 97.7$\%$ & 94.3$\%$ & 96.2$\%$ \\
		Quasar & 96.6$\%$ & 97.1$\%$ & 98.3$\%$ & 96.8$\%$ & 97.9$\%$ \\
		Star   & 93.7$\%$ & 95.4$\%$ & 96.8$\%$ & 94.5$\%$ & 96.4$\%$ \\
		\hline
	\end{tabular}
\end{table}
For the classification task, the following statistical metrics are used in this paper: precision, recall \citep{recall}, specificity, F1-score \citep{F1}, and accuracy, and the specific values are shown in Table~\ref{tab2}. Precision indicates the number of correctly classified positive samples as a proportion of the total number of samples predicted to be positive, and recall indicates the number of correctly classified positive samples as a proportion of the actual total number of positive samples.
The higher these two metrics are, the better, but they are a pair of contradictory metrics, so we use the F1-score (the summed average of precision and recall) to evaluate the classification results, and the formula is shown below.
\begin{equation}
	\mathrm{F} 1=\frac{2 \times \text { Precision } \times \text { Recall }}{\text { Precision }+\text { Recall }}
\end{equation}

\begin{figure}
	\centering
	\includegraphics[width=\linewidth]{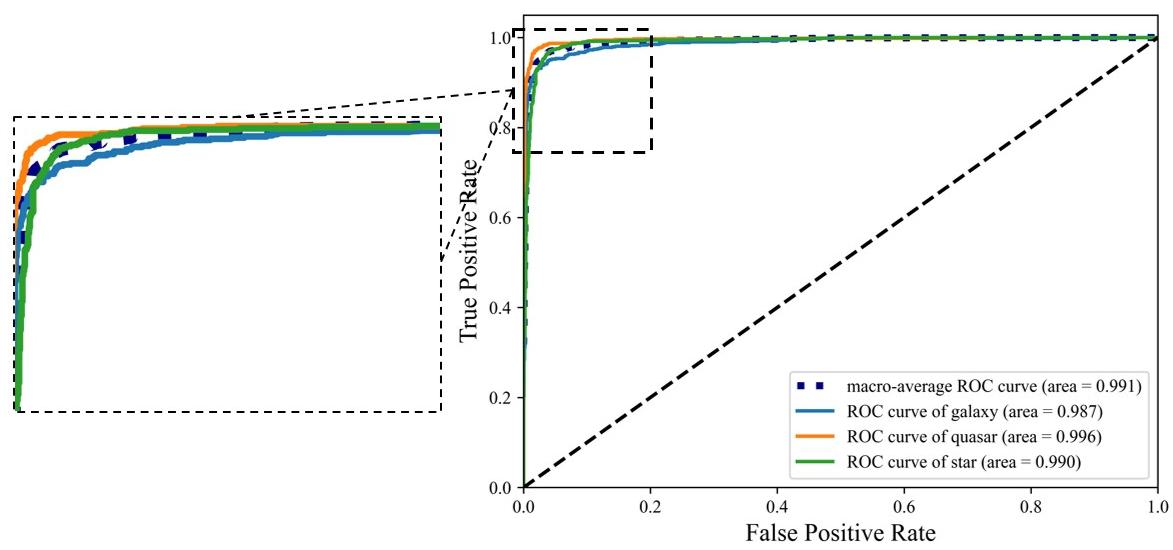}
	\caption{ROC for galaxies, quasars and stars.}
	\label{fig14}
\end{figure}

Specificity measures the classifier's ability to recognize positive examples; sensitivity measures the classifier's ability to recognize negative ones, which is calculated similarly to recall.
The Receiver Operating Characteristic curve (ROC) \citep{roc} can also prove the superiority of the classifier in this paper, as shown in Fig.~\ref{fig14}. The ROCs of galaxies, quasars, and stars all rise rapidly to around 1, effectively demonstrating that the algorithm in this paper has good classification results for all types of objects.

\begin{figure}
	\centering
	\subfloat[]{\includegraphics[scale=0.08]{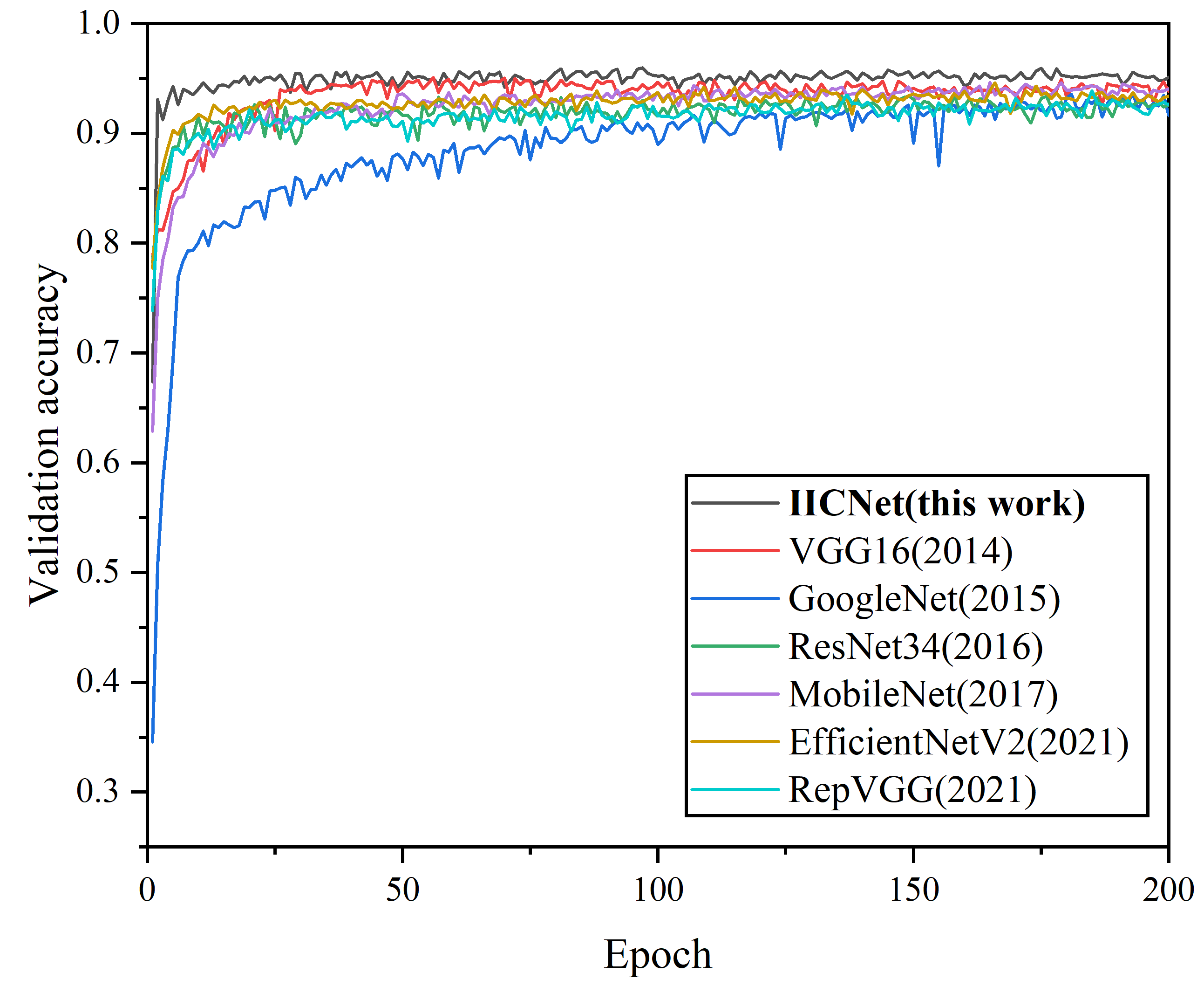}\label{fig15: a}}
	\quad
	\subfloat[]{\includegraphics[scale=0.08]{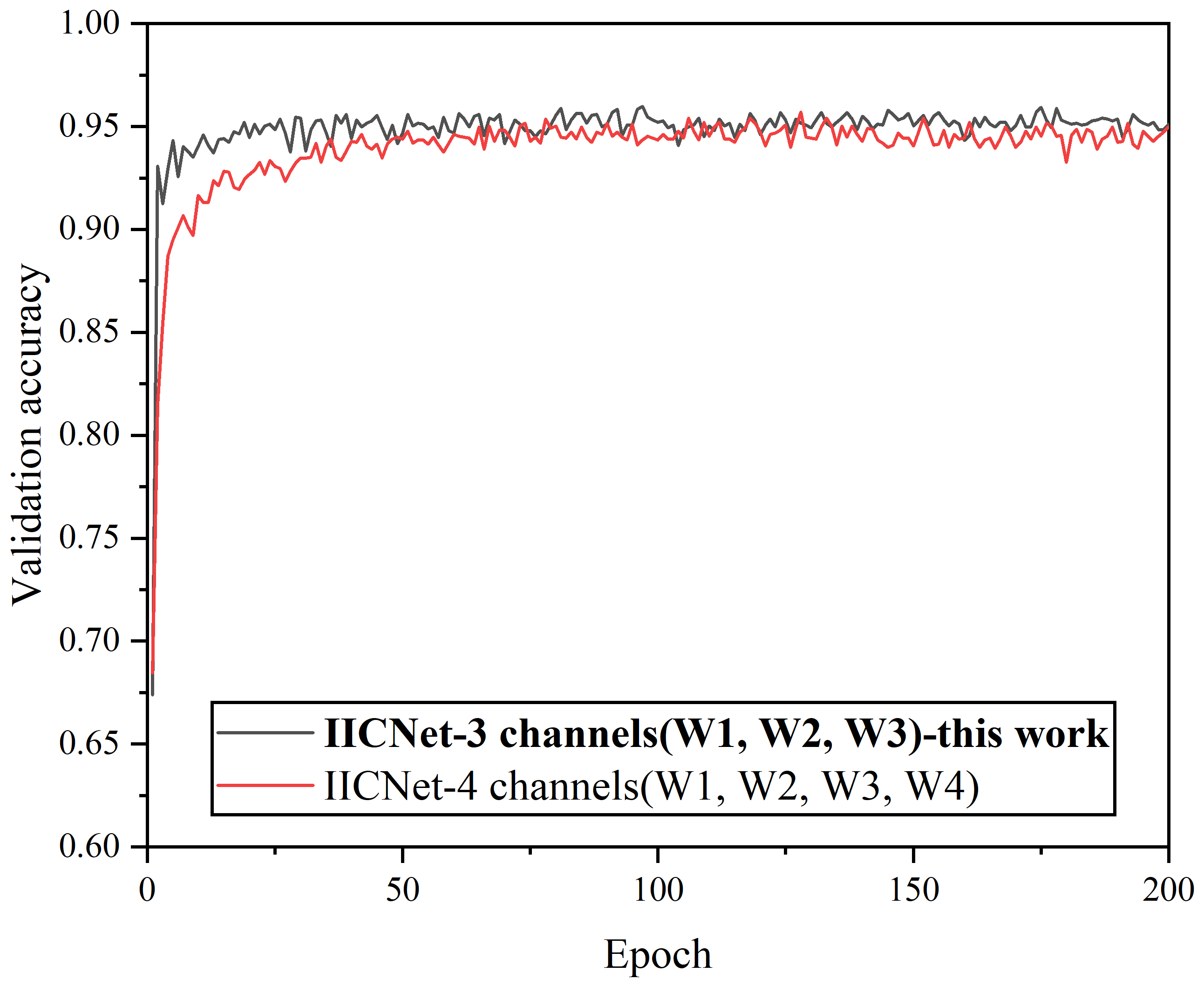}\label{fig15: b}}
	\caption{(a) Comparison results of IICnet and other image classification networks validation accuracy. (b) Comparison results of 3-channel (W1, W2, W3) and 4-channel (W1, W2, W3, W4) validation accuracy.}
	\label{fig15}
\end{figure}

\begin{figure}
	\centering
	\includegraphics[scale=0.05]{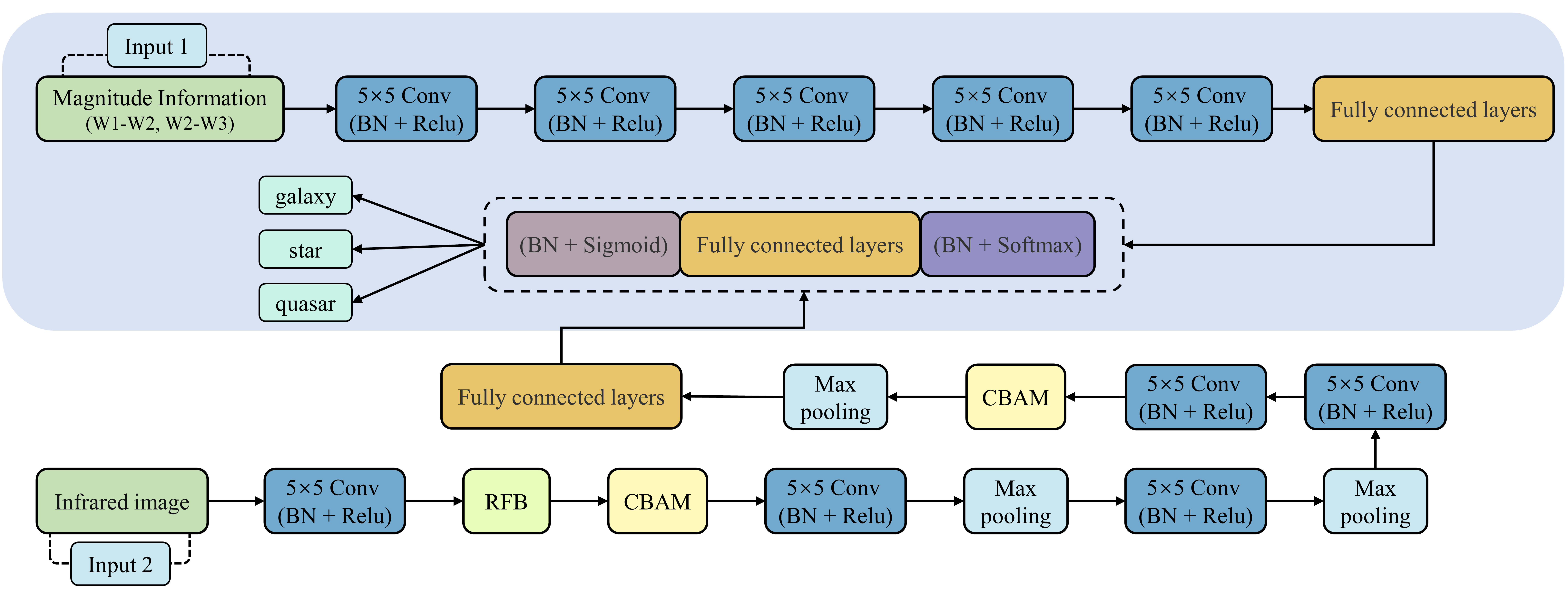}
	\caption{'Infrared image + color-color' classification network. The upper part, covered by the blue shade, is the color-color classification network.}
	\label{fig16}
\end{figure}

\subsection{Comparative Experiment}
\label{sec4.4}
This section compares IICnet with some classic novel classification networks, including VGG16 (2014) \citep{vgg}, GoogleNet (2015) \citep{google}, ResNet34 (2016) \citep{resnet}, Mobilenet (2017) \citep{Mobilenet}, EfficientNetV2 (2021) \citep{Efficientnetv}, and RepVGG (2021) \citep{repvgg} (EfficientNetV2 and RepVGG are the latest CNN-based networks we could find so far).
The accuracy curves on the validation set for each network are shown in Fig.~\ref{fig15: a}. Except for the comparison experiments using 7 models, this work is also experimented on different datasets (infrared images, spectra, color-color and 'infrared images + color-color').

It can be seem that the results of IICnet are better than the other mainstream classification networks. As shown in Fig.~\ref{fig15: a}, only IICnet can achieve more than 95$\%$ accuracy. Besides of this, it can maintain a small computational and parametric volume while improving accuracy, as shown in Table~\ref{tab3}. IICnet can obviously reduce the amount of computation by more than a half and the number of parameters by 1.47M compared to Mobilenet, which is the least computationally intensive way in Table~\ref{tab3}.

As mentioned in Section~\ref{sec2.2}, only W1, W2 and W3 bands are used to synthesize the images, due to the lower SNR of W4 band. The performance of using 3-channel and 4-channel images are conducted, which shows that the former are slightly better than the latter, as shown in Fig.~\ref{fig15: b}.

Color-color classification and 'infrared image + color-color' classification are based on revised IICnet, as shown in Fig.~\ref{fig16}, where the upper part covered with blue shading is the color-color classification network, and the composition of the upper and the lower form the 'infrared image + color-color' classification network.

\begin{figure}
	\centering
	\includegraphics[scale=0.11]{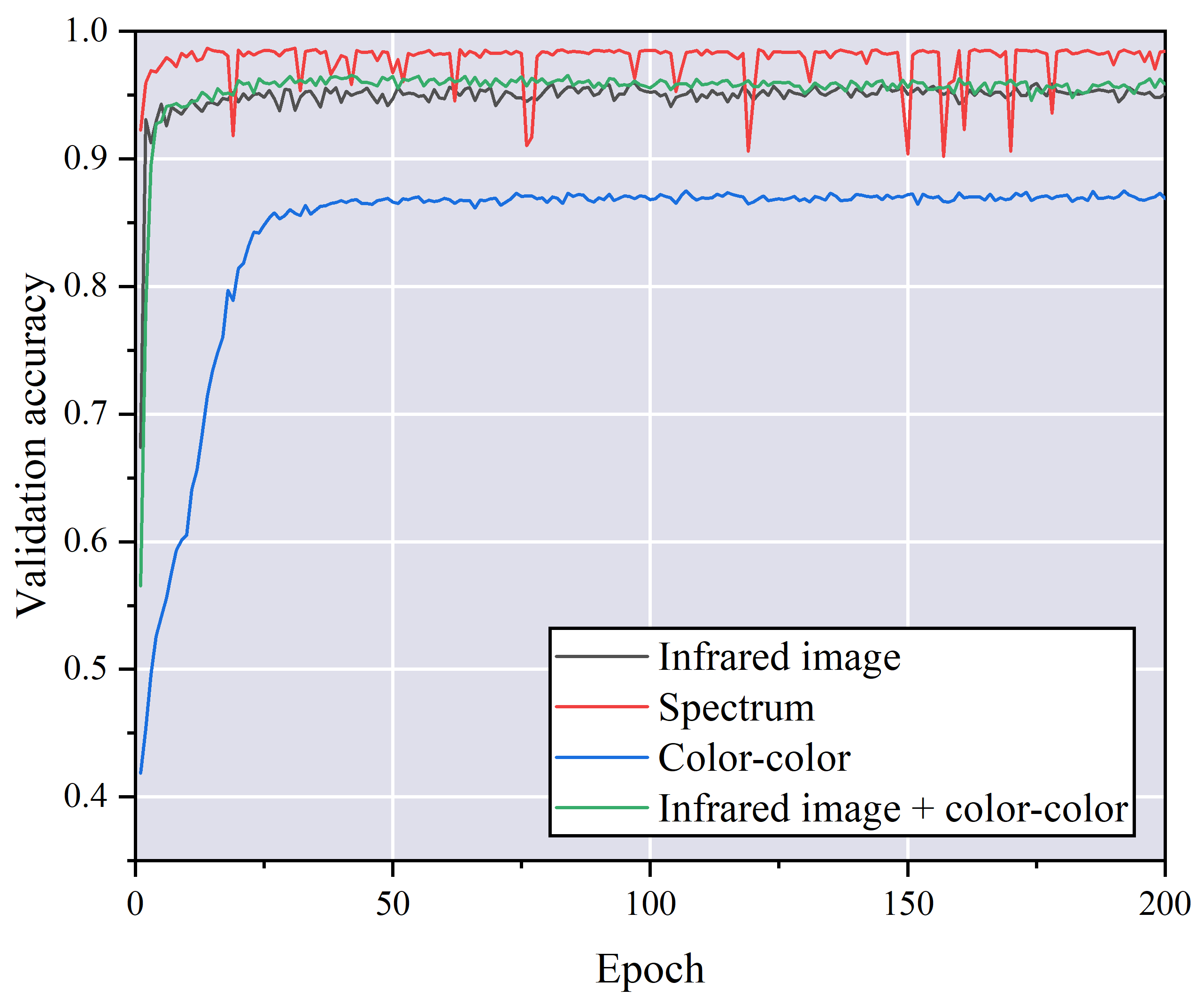}
	\caption{The accuracy curves of the validation sets corresponding to different data sets (infrared images, spectra, color-color, 'infrared images + color-color').}
	\label{fig17}
\end{figure}

The accuracy curves of the validation sets, which are resepctively obtained by IICnet and revised IICnet, are shown in Fig.~\ref{fig17}. Spectral classification has the highest accuracy, but it is difficult to obtain. The image classification accuracy can exceed 95$\%$, so using image classification will be a more common way. The color-color classification results are the worst, which also corresponds to the results shown in Fig.~\ref{fig6}. The results of 'infrared image + color-color classification' are about 1$\%$ higher than infrared image classification results. The reason is that some color information is lost when extracting features from infrared images, which can be alleviated by adding the magnitude information. The fused features will be further investigated in the subsequent work.

\begin{table}
	\centering
	\caption{Comparison of Flops and Params in the the seven networks.}
	\label{tab3}
	\begin{tabular}{lcc} 
		\hline
		Model & Flops & Params \\
		\hline
		VGG16 & 15.5G & 134.27M \\
		GoogleNet & 1.59G & 5.98M \\
		ResNet34   & 12.25G & 46.99M \\
		MobileNet & 587.94M & 3.22M \\
		EfficientNetV2 & 2.97G & 24.18M \\
		RepVGG & 3.83G & 30.07M \\
		\textbf{IICNet} & \textbf{218.66M} & \textbf{1.75M} \\
		\hline
	\end{tabular}
	\begin{tablenotes}
		\item\textit{Note}. Flops is the number of floating-point operations that can be used to measure the complexity of an algorithm. Params refers to the total number of parameters to be trained in the model training, which is used to measure the size of the model (computational complexity).
	\end{tablenotes}
\end{table}

\begin{figure}
	\centering
	\includegraphics[scale=0.08]{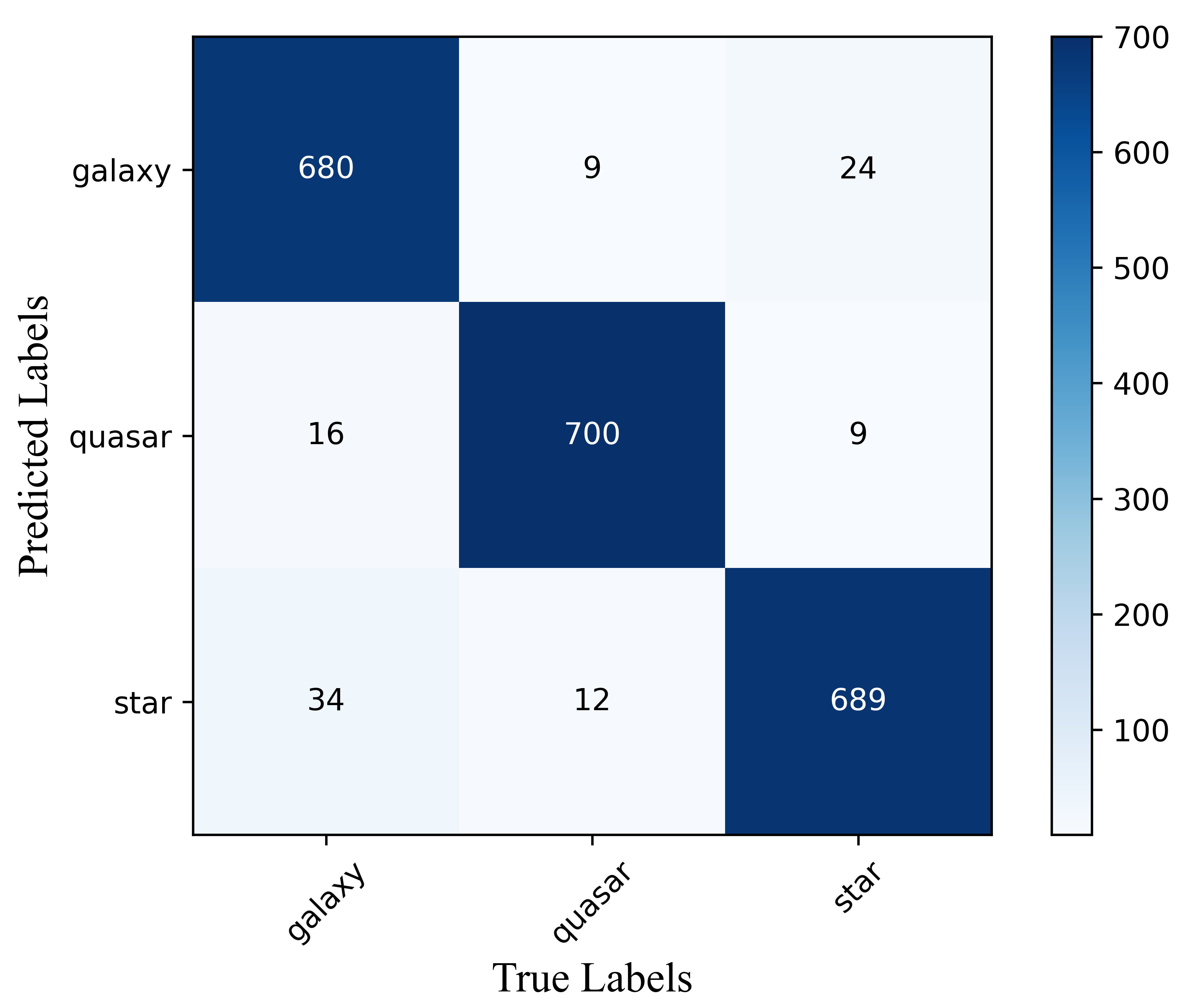}
	\caption{Confusion Matrix of IICnet. Each column of the confusion matrix represents the number of true labels for each class, and each row represents the number of predicted labels for each class.}
	\label{fig18}
\end{figure}

\begin{table}
	\centering
	\caption{The confidence of the three samples in Fig.~\ref{fig4}.}
	\label{tab4}
	\begin{threeparttable}
		\begin{tabular}{lcccc} 
			\hline
			& & & Classification probability & \\
			Class & Sample & galaxy & star & quasar \\
			\hline
			galaxy & \raisebox{-.5\height}{\includegraphics[scale=1]{f4a.jpg}} & \textbf{0.999486} & 0.000453 & 0.000061 \\
			star & \raisebox{-.5\height}{\includegraphics[scale=1]{f4b.jpg}} & 0.000157 & \textbf{0.999814} & 0.000029 \\
			quasar & \raisebox{-.5\height}{\includegraphics[scale=1]{f4c.jpg}} & 0.024439 & 0.000004 &  \textbf{0.975557} \\
			\hline
		\end{tabular}
		\begin{tablenotes}
			\item \textit{Note}. The confidence that each image falls into three types can be obtained, after the three images are input into IICnet.
		\end{tablenotes}
	\end{threeparttable}
\end{table}

\begin{figure}
	\centering
	\includegraphics[width=\linewidth]{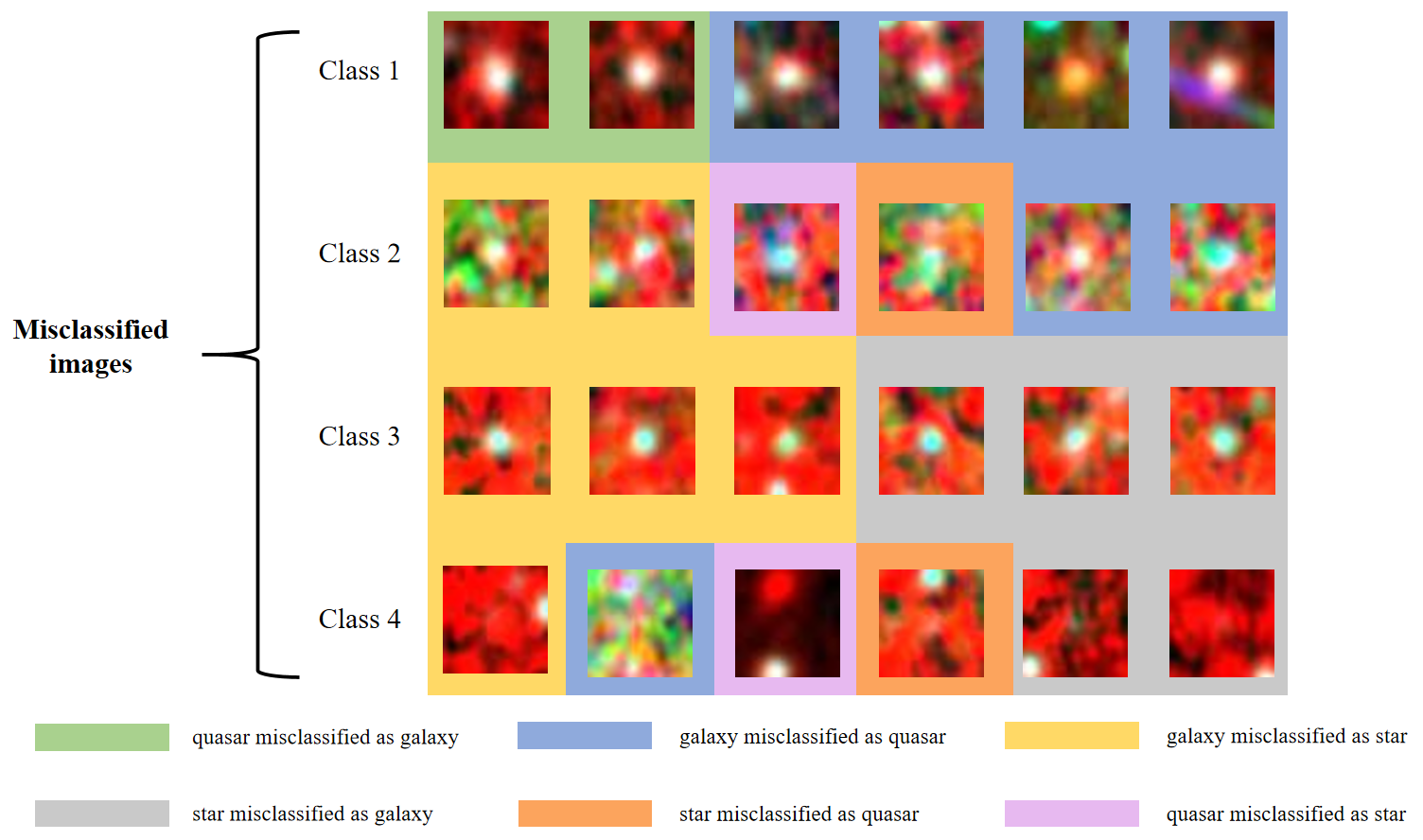}
	\caption{A few misclassified images. Class 1, 2, and 3 are the three types obtained by K-means. Class 4 involves some images in which the source is obscured or absent entirely from the center.}
	\label{fig19}
\end{figure}

\subsection{Confusion Matrix}
\label{sec4.5}
The confusion matrix can be used to demonstrate the classification effect. The confusion matrix drawn for the test set in this paper is shown in Fig.~\ref{fig18}. Of these, the number of misclassified samples is tiny, with the vast majority concentrated on the diagonal.

The histogram in Section~\ref{sec2.2} (Fig.~\ref{fig5}) cannot distinguish the types to which the three images in Fig.~\ref{fig4}, but inputting the three images into the IICnet model gives evident confidence in the classification, as shown in Table~\ref{tab4}. All three images are classified correctly, with a confidence level close to 1.

\section{DISCUSSION}
\label{sec5}
\subsection{Analysis of misclassified samples}
\label{sec5.1}
In Fig.~\ref{fig18}, there are 104 misclassified images, which are divided into 4 classes, namely Class 1 (37 images), Class 2 (13 images), Class 3 (45 images), and Class 4 (9 images). Some examples of misclassified images are shown in Fig.~\ref{fig19}, and the analysis is as follows.

K-means clustered the misclassified samples to obtain three classes of images: Class 1, Class 2, and Class 3. Visually, it can be seen that the images in Class 1 are darker, mainly showing the confusion of galaxies and quasars; in Class 2, the colors are complex, so the misclassification is more complicated; and in Class 3, the colors are brighter, mainly showing the confusion of galaxies and stars. How to further distinguish these images requires more effort in future work. Class 4 is a particular type found in misclassified samples, because its sample center has no source, which is unfavorable for feature extraction in IICnet. IICnet is more concerned with central sources, as evidenced by Section~\ref{sec3.2}, so how to handle such images is to be considered in the subsequent work.

\begin{figure}
	\centering
	\includegraphics[width=\linewidth]{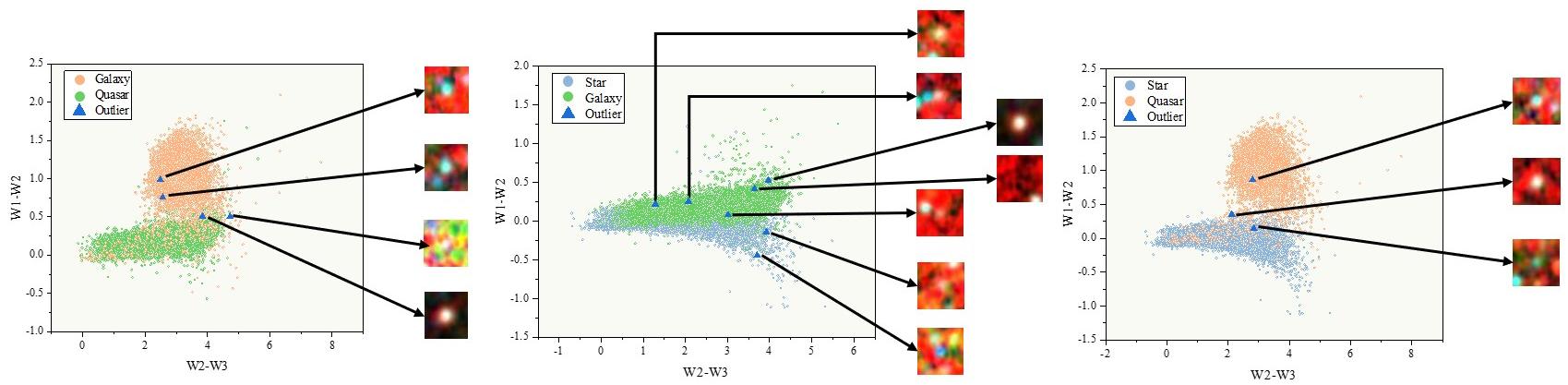}
	\caption{Samples with correct classification but low confidence. The blue triangle represents the outlier sample.}
	\label{fig20}
\end{figure}

\subsection{Analysis of outlier samples}
\label{sec5.2}
In addition to the misclassified samples, some images are correctly classified but have low confidence in the classification, which are called outlier samples in this paper. These samples have features easily confused with other types, so it is necessary to analyze them.

\begin{figure}
	\centering
	\includegraphics[width=\linewidth]{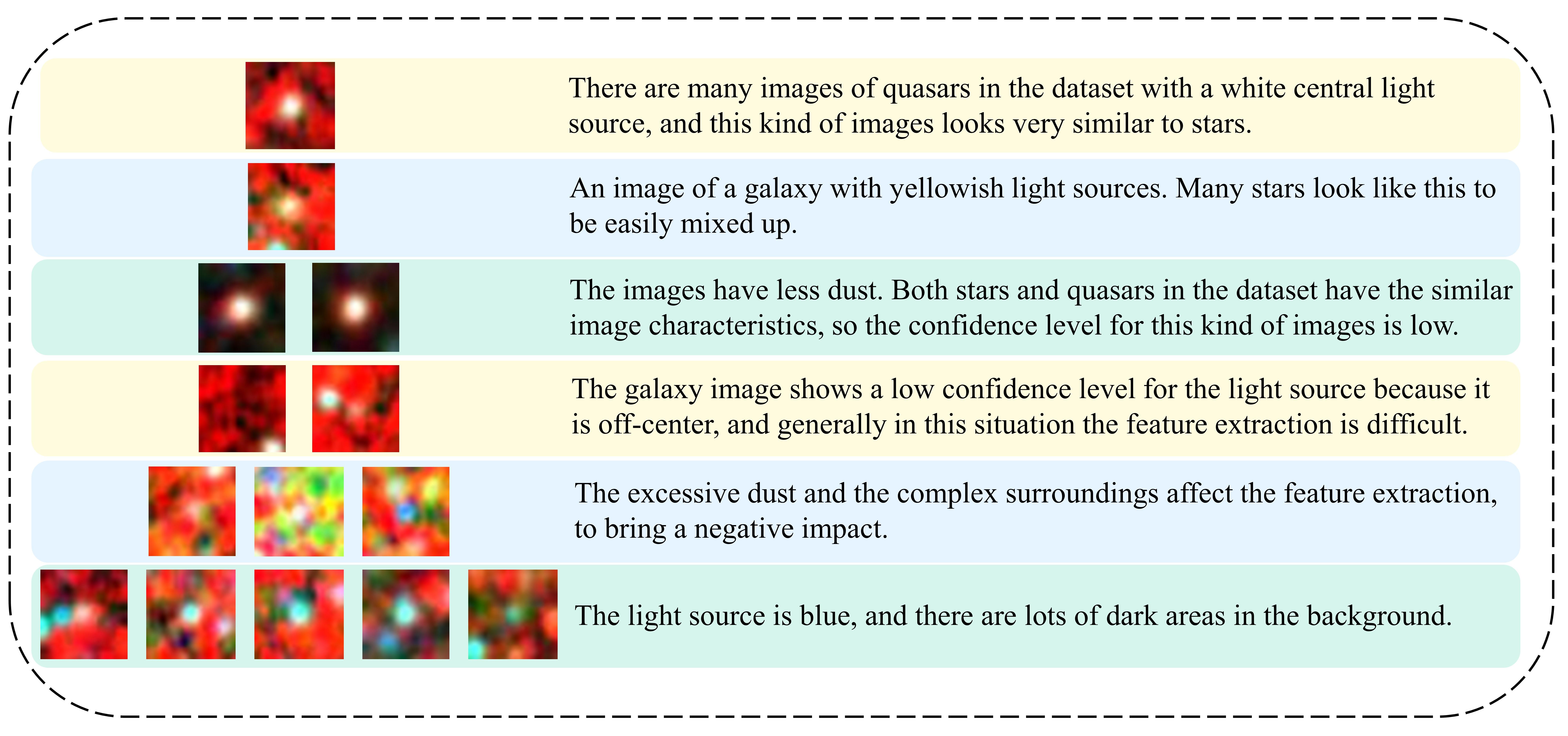}
	\caption{Analysis of outlier samples. Manually watching and labeling the outlier samples, and giving the characteristics of the images based on color distribution and textures, and summarizing the images’ characteristics for each case.}
	\label{fig21}
\end{figure}

When the test set is inputted into IICnet for testing, the classification confidence for each image is obtained. By filtering the classification confidence, the filtering condition is the images with a confidence below 0.6, although the classification is correct. A total of 14 images were chosen, as shown in Fig.~\ref{fig20}, and combined with Figure 6 to facilitate viewing the distribution. According to the image characteristics, the analysis of these samples are presented in Fig.~\ref{fig21} and are divided into six cases. Each case has its unique characteristics. The classifier in this paper gets a lower confidence level in distinguishing images whose features need to be clarified but still gets correct classification results. The above proves the superiority of IICnet.

\section{CONCLUSIONS}
\label{sec6}
The task of the infrared image classification of galaxies, quasars, and stars has been rarely reported in past literatures. And for many images it becomes extremely difficult owing to the complexity of the images and similarities between different types. This paper uses W1, W2, and W3 for WISE to synthesize RGB images and specifically designs the IICnet to classify infrared images into galaxies, quasars, and stars. IICnet intergrates RFB and CBAM (Section~\ref{sec3.1}), which improve feature extraction for the sources and enable higher classification accuracy rates. In the experiments, by comparing IICnet with VGG16, GoogleNet, Resnet34, MobileNet, EfficientNetV2, and RepVGG, it is proved that IICnet outperforms all the other methods for the classification of infrared images.

For the analysis of misclassification samples, K-means clustering is used and 4 cases are discussed. Case 1, 2, 3 are misclassified because the images' features are highly similar. Case 4 misclassified because the source is off-center and cannot be extracted efficiently.

Outliers are also analyzed which are the correctly classified images but with low confidence. Outliers are at the borders of the types.  Because the confidence level is low, it seems to be lucky that they can be classified correctly by the current method, IICnet.  In the future, support vector machine (SVM) mechanism may be considered to be used because the ourliers here are like support vectors.

In summary, experiments have proven that IICnet is very effective in classifying infrared images. It may provide a new tool for astronomers.  Of course it can be futhur enhanced by a better feature extraction block, a new post-processing block like SVM, etc.
%

\begin{acknowledgements}

This work is supported by the Natural Science Foundation of Tianjin (22JCYBJC00410) and the Joint Research Fund in Astronomy, National Natural Science Foundation of China (U1931134). We are grateful for Sloan Digital Sky Survey (SDSS) and Wide-field Infrared Survey Explorer (WISE) that provide us with open data.

\end{acknowledgements}

\bibliographystyle{raa}
\bibliography{bibtex}

\begin{thebibliography}{35}
\providecommand\natexlab[1]{#1}
\providecommand\JournalTitle[1]{#1}

\bibitem[Bai {et~al.}(2018)]{Bai}
Bai, Y., Liu, J., Wang, S., \& Yang, F. 2018, \aj, 157, 9

\bibitem[Chawla {et~al.}(2002)]{roc}
Chawla, N.~V., Bowyer, K.~W., Hall, L.~O., \& Kegelmeyer, W.~P. 2002, Journal
  of artificial intelligence research, 16, 321

\bibitem[Chinchor \& Sundheim(1993)]{F1}
Chinchor, N., \& Sundheim, B.~M. 1993, in Fifth Message Understanding
  Conference (MUC-5): Proceedings of a Conference Held in Baltimore, Maryland,
  August 25-27, 1993

\bibitem[Cutri {et~al.}(2013)]{Cutri}
Cutri, R., Wright, E., Conrow, T., {et~al.} 2013, Explanatory Supplement to the
  AllWISE Data Release Products, 1

\bibitem[Ding {et~al.}(2021)]{repvgg}
Ding, X., Zhang, X., Ma, N., {et~al.} 2021, in Proceedings of the IEEE/CVF
  Conference on Computer Vision and Pattern Recognition, 13733

\bibitem[Duxbury \& Soifer(1980)]{Duxbury}
Duxbury, J.~H., \& Soifer, B. 1980, in Applications of Digital Image Processing
  to Astronomy, Vol. 264, SPIE, 81

\bibitem[Egmont-Petersen {et~al.}(2002)]{Egmont}
Egmont-Petersen, M., de~Ridder, D., \& Handels, H. 2002, Pattern recognition,
  35, 2279

\bibitem[Erickson(1992)]{EricksonEF}
Erickson, E. 1992, Next Generation Infrared Space Observatory, 61

\bibitem[Erickson {et~al.}(1985)]{Erickson}
Erickson, E., Houck, J., Harwit, M., {et~al.} 1985, Infrared Physics, 25, 513

\bibitem[Ethiraj \& Bolla(2022)]{ethiraj2022classification}
Ethiraj, S., \& Bolla, B.~K. 2022, arXiv preprint arXiv:2205.10745

\bibitem[Glass(1999)]{Glass}
Glass, I.~S. 1999, Handbook of infrared astronomy No.~1 (Cambridge University
  Press)

\bibitem[Guo {et~al.}(2022)]{GuoXiaoyu}
Guo, X., Liu, C., Qiu, B., {et~al.} 2022, \mnras, 517, 1837

\bibitem[Harrington(2012)]{recall}
Harrington, P. 2012, Machine learning in action (Simon and Schuster)

\bibitem[He {et~al.}(2016)]{resnet}
He, K., Zhang, X., Ren, S., \& Sun, J. 2016, in Proceedings of the IEEE
  conference on computer vision and pattern recognition, 770

\bibitem[He {et~al.}(2021)]{HeZhendong}
He, Z., Qiu, B., Luo, A.-L., {et~al.} 2021, \mnras, 508, 2039

\bibitem[Howard {et~al.}(2017)]{Mobilenet}
Howard, A.~G., Zhu, M., Chen, B., {et~al.} 2017, arXiv preprint
  arXiv:1704.04861

\bibitem[Kessler {et~al.}(1996)]{Kessler}
Kessler, M., Steinz, J., Anderegg, M., {et~al.} 1996, A\&A, 315, L27

\bibitem[Kim \& Brunner(2016)]{kim2016star}
Kim, E.~J., \& Brunner, R.~J. 2016, Monthly Notices of the Royal Astronomical
  Society, stw2672

\bibitem[Kingma \& Ba(2014)]{Adam}
Kingma, D.~P., \& Ba, J. 2014, arXiv preprint arXiv:1412.6980

\bibitem[Kurcz {et~al.}(2016)]{Kurcz}
Kurcz, A., Bilicki, M., Solarz, A., {et~al.} 2016, A\&A, 592, A25

\bibitem[Lintott {et~al.}(2008)]{lintott2008galaxy}
Lintott, C.~J., Schawinski, K., Slosar, A., {et~al.} 2008, \mnras, 389, 1179

\bibitem[Liou(2002)]{Liou}
Liou, K.-N. 2002, An introduction to atmospheric radiation, Vol.~84 (Elsevier)

\bibitem[Liu {et~al.}(2018)]{RFB}
Liu, S., Huang, D., {et~al.} 2018, in Proceedings of the European conference on
  computer vision (ECCV), 385

\bibitem[Liu {et~al.}(2016)]{softmax}
Liu, W., Wen, Y., Yu, Z., \& Yang, M. 2016, arXiv preprint arXiv:1612.02295

\bibitem[Martinazzo {et~al.}(2021)]{Martinazzo}
Martinazzo, A., Espadoto, M., \& Hirata, N.~S. 2021, in 2020 25th International
  Conference on Pattern Recognition (ICPR), IEEE, 4176

\bibitem[Peng {et~al.}(2017)]{largekernel}
Peng, C., Zhang, X., Yu, G., Luo, G., \& Sun, J. 2017, in Proceedings of the
  IEEE conference on computer vision and pattern recognition, 4353

\bibitem[Simonyan \& Zisserman(2014)]{vgg}
Simonyan, K., \& Zisserman, A. 2014, arXiv preprint arXiv:1409.1556

\bibitem[Steinwart \& Christmann(2008)]{Steinwart}
Steinwart, I., \& Christmann, A. 2008, Support vector machines (Springer
  Science \& Business Media)

\bibitem[Szegedy {et~al.}(2015)]{google}
Szegedy, C., Liu, W., Jia, Y., {et~al.} 2015, in Proceedings of the IEEE
  conference on computer vision and pattern recognition, 1

\bibitem[Tan \& Le(2021)]{Efficientnetv}
Tan, M., \& Le, Q. 2021, in International Conference on Machine Learning, PMLR,
  10096

\bibitem[Touvron {et~al.}(2019)]{imagesize}
Touvron, H., Vedaldi, A., Douze, M., \& J{\'e}gou, H. 2019, Advances in neural
  information processing systems, 32

\bibitem[Tu \& Wang(2013)]{Tu}
Tu, X., \& Wang, Z.-X. 2013, \raa, 13, 323

\bibitem[Wandell \& Winawer(2015)]{pRF}
Wandell, B.~A., \& Winawer, J. 2015, Trends in cognitive sciences, 19, 349

\bibitem[Woo {et~al.}(2018)]{CBAM}
Woo, S., Park, J., Lee, J.-Y., \& Kweon, I.~S. 2018, in Proceedings of the
  European conference on computer vision (ECCV), 3

\bibitem[Wright {et~al.}(2010)]{Wright}
Wright, E.~L., Eisenhardt, P.~R., Mainzer, A.~K., {et~al.} 2010, \aj, 140, 1868

\end{thebibliography}
\label{lastpage}

\end{document}